\documentclass{amsart}

\usepackage{epsfig}
\usepackage{amssymb}

\usepackage{amscd}
\usepackage{graphics}
\newtheorem{Theorem}{Theorem}[section]
\newtheorem{Proposition}{Proposition}[section]
\newtheorem{Corollary}{Corollary}[section]
\newtheorem{Lemma}{Lemma}[section]

\def\proof{\par{\it Proof}. \ignorespaces}
\def\endproof{{\ \vbox{\hrule\hbox{%
   \vrule height1.3ex\hskip0.8ex\vrule}\hrule }}\par}
\newenvironment{Proof}{\proof}{\endproof}

\theoremstyle{definition}
\newtheorem{Definition}[Theorem]{Definition}

\theoremstyle{remark}
\newtheorem{Remark}[Theorem]{Remark}

\numberwithin{equation}{section}

\begin{document}

\renewcommand\baselinestretch{1.2}


\title{NORMAL FORM AND SOLITONS}

\author{Yasuaki Hiraoka}
\address{Department of Informatics and Mathematical Science, Osaka University, Toyonaka, Osaka 562-
Japan}
\email{}
\thanks{The second author is supported in part by NSF Grant \#DMS0071523.}

\author{Yuji Kodama}
\address{Department of Mathematics, Ohio State University, Columbus,
OH 43210}
\email{kodama@math.ohio-state.edu}

\newcommand{\bsquare}{\hbox{\rule{6pt}{6pt}}}

\begin{abstract}
We present a review of the normal form theory for weakly dispersive nonlinear wave equations where the leading order phenomena can be described by the KdV equation. This is an infinite dimensional extension of the well-known Poincar\'e-Dulac normal form theory for ordinary differential equations. We also provide a detailed analysis of the interaction problem of solitary waves
as an important application of the normal form theory. Several explicit examples are discussed based on the normal form theory, and the results are compared with their numerical simulations. Those examples include the ion acoustic wave equation, the Boussinesq equation as a model of the shallow water waves, the regularized long wave equation and the Hirota bilinear equation having a 7th order linear dispersion. 
\end{abstract}

\maketitle
\markboth{HIRAOKA AND KODAMA}
  {NORMAL FORM AND SOLITONS}

\thispagestyle{empty}
\pagenumbering{roman}
\tableofcontents

\pagenumbering{arabic}
\setcounter{page}{1}
\section{Introduction}
In this chapter, we review the normal form theory developed in
\cite{kodama:85, bbmnormal} for weakly nonlinear and weakly dispersive wave
equations where the leading order equation is given by the KdV
equation in an asymptotic perturbation sense. The chapter is based on the
report \cite{kdvnormal} and the master thesis of the first author at Osaka
University.

A series of lectures was carried out by the second author in the Euro Summer
School 2001 held at the Isaac Newton Institute, Cambridge for
August 12-25, 2002. The lecture started with a brief summary of the
Poincar\`e-Dulac normal form theory for a system of ordinary differential
equations \cite{arnold}.  The main point in the lecture is to present the
normal form theory for {\it near} integrable system where the leading order
system is given by a nolinear wave equation.
This may be considered as an infinite dimensional extension of the
Poincar\'e-Dulac normal form theory. The basic technique of the normal form
theory based on the Lie transformation can be formally extended to the
infinite dimensional case. Unfortunately the convergence theorem of the
normal form series has no extension to the present theory.
However one should emphasize that the leading order in the present theory is
given by a nonlinear equation, and it is not clear how one define a resonant
surface for the leading order equation. This may provide a good problem for
a future project.

Let us briefly summarize a background of the normal form theory for near
integrable systems of nonlinear dispersive equations.
It is well-known that for a wide class of nonlinear dispersive wave
equations, the leading order nonlinear equation in an asymptotic expansion
turns out to be given by an integrable system, such as the Korteweg-de
Vries (KdV) equation in weak dispersion limit and the nonlinear
Schr\"odinger equation in strong dispersion limit (see for example
\cite{taniuti}). This implies that most of the nonlinear dispersive wave
equations are integrable at the nontrivial leading orders in an asymptotic
sense. Then a natural question is to ask how the higher order corrections
affect to the integrability of the leading order equations. In
\cite{kodamataniuti}, the effect of the higher order corrections on
one-soliton solution of the KdV equation was studied, and it was shown that
the velocity of soliton is shifted by the secular terms in the higher order
terms. Those secular terms or resonant terms are given by the symmetries of
the KdV equation. The nonsecular terms then contribute to modify the shape
of soliton.
However the multi-soliton interactions were not studied in that paper.
In \cite{kodama:85}, the normal form for weakly dispersive equations was
first introduced up to the second oreder corrections, and it was found that
the integrable approximation can be extended beyond the KdV approximation
but not to the second order. This obstacle to the {\it asymptotic}
integrability plays no rule for one-soliton solution, but provides a crucial
effect for two soliton interactions. This was found in
\cite{bbmnormal, kdvnormal}. The obstacles are defined as the nonexistence of
the integrals of perturbed equation in the form of the power series in a
small parameter. This was also recognized as the nonexistence of approximate
symmetries of the perturbed equation \cite{mikhailov:91}.
(The normal form for strongly dispersive wave equations has been also
studied in \cite{kano:89, obstacle}.) Then in \cite{bbmnormal}, the effect
of the obstacle on the interaction of two soliary waves was studied for the
regularized long wave equation (although the method can apply to other
equations of weakly dispersive system). An inelasticity due to the obstacle
was found, and it leads to the shifts of the soliton parameters and a
generation of a new soliton as well as radiations through the interaction.

In this lecture, we present a comprehensive study of the normal form theory
for weakly dispersive wave equations: We start in Section 2 to define the
perturbed KdV equation as an asymptotic expansion of a weakly dispersive
wave equation whose leading order equation is given by the KdV equation. We
give a recursion formula to generate the higher order corrections which may
be obtained by an asymptotic perturbation method (see for example
\cite{taniuti}). The set of those higher order terms forms an extended space
of differential polynomials which includes some nonlocal terms. The space is
denoted by $\widehat{\mathcal P}_{{\rm odd}}$, where \lq\lq odd\rq\rq  implies
the weight of the polynomial.

In Section 3, the conserved quantities or integrals of the KdV equation are
reviewed, and we discussed approximate integrals of the perturbed KdV
equation. We then obtain the conditions for the existence of approximate
integrals in each order (Proposition \ref{PRO:conservation}). We also
discuss a connection of the conserved quantities and the $N$-soliton
solutions of the KdV equation.

In Section 4, we review the symmetries of the KdV equation, and discuss the
approximate symmetries for the perturbed KdV equation.
Here we also define the space of $\widehat{\mathcal P}_{{\rm even}}$, which
together with $\widehat{\mathcal P}_{{\rm odd}}$ provides the appropriate
spaces for the normal form transform defined in the next section.

In Section 5, we describe the normal form theory. The normal form
transformation is then obtained by a linear equation of an adjoint map
defined as
\[
{\rm ad}_{K^{(0)}}:\widehat{\mathcal P}_{{\rm even}}\longrightarrow
\widehat{\mathcal P}_{{\rm odd}},\]
where $K^{(0)}$ is the KdV vector field. The explicit form of the normal
form is given for the perturbed KdV equation which contains the first three
lowest weight approximate conserved quantities (Proposition 5.1). The normal
form then admits one-soliton solution of the KdV equation, which confirms
the result in \cite{kodamataniuti}.
We also discuss the Gardner-Miura transformation which is an invertible
version of the Miura transformation, and show that the inverse Gardner-Miura
transformation is nothing but the normal form transformation after removing
the symmetries of the KdV equation (Proposition 5.2).

In Section 6, we consider the interaction problem of two solitary waves, and
provides the explicit formulae for the shifts of the soliton parameters
(Proposition 6.1). we also give the formulae of the radiation energy and
additional phase shifts which are used to compare with the numerical
simulations for some examples in the next section.

Finally in Section 7, we present explicit examples including ion acoustic
wave equation, the Boussinesq equation as a model of shallow water waves and the
regularized long wave equation. We show the good agreements with the results
obtained by the normal form theory.
We also consider a 7th order Hirota bilinear equation which admits an exact
two soliton solution but is known to be nonintegrable.
We look for an obstacle to the asymptotic integrability, and find that the
obstacles appear at the fourth order. This implies that the forth order
obstacles play no rule for two soliton solution, just like all obstacles has
no rule for one-soliton solution for the system with the first three
approximate integrals.

\section{Perturbed KdV equation}
Under the assumption of weak nonlinearity and weak dispersion, the wave propagation in a one-dimensional medium can be described by the KdV equation in the leading order of
an asymptotic expansion. Using an appropriate asymptotic perturbation method
(e.g. see \cite{taniuti}) one can show that the higher order correction to the KdV equation has the following expansion form with a small
parameter $\epsilon$ with $0<\epsilon \ll 1$,
\begin{eqnarray}
&&u_{t}+K(u;\epsilon)=O(\epsilon^{N+1}), \label{eq:pkdv}\\
&&\quad {\rm with}\quad K(u;\epsilon)=K^{(0)}(u)+\epsilon K^{(1)}(u)+\epsilon^2 K^{(2)}(u)+\cdots + \epsilon^{N} K^{(N)}(u), 
\nonumber
\end{eqnarray} 
where $K^{(0)}(u)$ gives the KdV flow and the higher order corrections $K^{(n)}(u)$ are generated by a recursion formula starting from $n=-1$,
\begin{equation}
\label{perturbation}
\begin{array}{ll}
K^{(n)}(u) &=\displaystyle{\sum_{i=1}^{M(n)}a_i^{(n)}X_i^{(n)}(u)} \\
&=a_1^{(n)}u_{(2n+3)x} 
 +\displaystyle{\sum_{n_1+n_2 =n-2 \atop{
1\le i\le M(n_1) \atop 1\le j\le M(n_2)}} c_{ij}^{(n)}\left(X_i^{(n_1)}
D^{-1}X_j^{(n_2)}\right)(u)}.
\end{array}
\end{equation}
Here $a_i^{(n)},~ c_{ij}^{(n)}$ are the real constants determined by the original physical problem, $u_{nx}=\partial^n u/\partial x^n$, and $D^{-1}$ indicates an 
integral over $x$, $D^{-1}(\cdot):=\int_{-\infty}^x dx'(\cdot)$. Each $X_i^{(n)}(u)$
is a monomial in the polynomial $K^{(n)}(u)$, and $M(n)$ is the total number
of independent monomials of the order $n$. The first few terms of $K^{(n)}(u)$ are then given by
\[
\begin{array}{llll}
K^{(-1)}&=&a_1^{(-1)}u_x, &M(-1)=1, \\
K^{(0)}&=&a_1^{(0)}u_{3x}+a_2^{(0)}uu_{x}, &M(0)=2, \\
K^{(1)}&=&a^{(1)}_{1}u_{5x}+a^{(1)}_{2}u_{3x}u+a^{(1)}_{3}u_{2x}u_{x}+
a^{(1)}_{4}u_{x}u^2, & M(1)=4, \\
K^{(2)}&=&a^{(2)}_{1}\!u_{7x}\!+\!a^{(2)}_{2}\!u_{5x}\!u\!+\!a^{(2)}_{3}\!u_{4x}\!u_{x}\!+\!
a^{(2)}_{4}\!u_{3x}\!u^{2}\!+\!a^{(2)}_{5}\!u_{3x}\!u_{2x}\!  \\
&{}& +a^{(2)}_{6}\!u_{2x}\!u_{x}\!u\!+a^{(2)}_{7}\!u_{x}\!u^{3}\!+\!a^{(2)}_{8}\!u^{3}_{x},  & M(2)=8, \\
K^{(3)}&=& a^{(3)}_{1}u_{9x}+a^{(3)}_{2}u_{7x}u+a^{(3)}_{3}u_{6x}u_{x}+
a^{(3)}_{4}u_{5x}u_{2x}\\
&{}& +a^{(3)}_{5}u_{5x}u^{2}+a^{(3)}_{6}u_{4x}u_{3x}+a^{(3)}_{7}u_{4x}u_{x}u +a^{(3)}_{8}u_{3x}u_{2x}u \\
&{}&+a^{(3)}_{9}u_{3x}u^{2}_{x}
+a^{(3)}_{10}u_{3x}u^{3}+a^{(3)}_{11}u^{2}_{2x}u_{x} +a^{(3)}_{12}u_{2x}u_{x}u^2\\
&{}& +a^{(3)}_{13}u^{3}_{x}u +a^{(3)}_{14}u_{x}u^{4}+a^{(3)}_{15}u_{x}D^{-1}(u^{3}_{x}), &M(3)=15.
\end{array}\]
We normalize the KdV flow so that $K^{(0)}(u)$ is given by the standard form which we denote by $K^{(0)}_0(u)$, i.e.
\[ K_0^{(0)}=u_{3x} + 6uu_x.\]
Each polynomial $K^{(n)}$ has the scaling property: Assign the weight 2 to $u(x,\cdot)$,
and 1 to $\partial /\partial x$. Then if $u(x,\cdot)=\delta^2v(\delta x,\cdot)=\delta^2v(\xi,\cdot)$,
we have
\[K^{(n)}(u(x,\cdot))=\delta^{2n+5}K^{(n)}(v(\xi,\cdot))=
\delta^{2n+5}(a_1^{(n)}v_{(2n+5)\xi}+\cdots).\]
Thus each polynomial $K^{(n)}(u)$ has the homogeneous weight \lq\lq $2n+5$\rq\rq.
We denote by $\widehat{\mathcal P}_{{\rm odd}}[u]$ the set of all the odd weight polynomials generated by
the formula (\ref{perturbation}). Then we have
\[
\widehat{\mathcal P}_{{\rm odd}}[u]=\bigoplus_{n=-1}^{\infty}\widehat{\mathcal P}_{{\rm odd}}^{(n)}[u],
\]
where  $\widehat{\mathcal P}_{{\rm odd}}^{(n)}[u]$ is 
the finite dimensional subspace of the polynomials with the homogeneous weight $2n+5$, and
the dimension of the space is given by ${\rm dim}~\widehat{\mathcal P}_{{\rm odd}}^{(n)}[u]=M(n)$, 
\[
\widehat{\mathcal P}_{{\rm odd}}^{(n)}[u]={\rm Span}_{\mathbb R}\left\{~ X_i^{(n)}(u)~:~ 1\le i\le M(n)~\right\}.
\]
As shown in the examples, the subspaces $\widehat{\mathcal P}_{{\rm odd}}^{(n)}[u]$
up to $n=2$ are given by the differential polynomial of $u$ and its derivatives,
and we denote
\[
\widehat{\mathcal P}_{{\rm odd}}^{(n)}[u]={\mathcal P}_{2n+5}[u]\oplus_{\mathbb R}
{\mathcal Q}_{2n+5}[u],\]
where ${\mathcal P}_k[u]$ is the space of homogeneous differential polynomials of weight $k$,
\[
{\mathcal P}_k[u]={\rm Span}_{\mathbb R}\left\{~u^{l_0}u_x^{l_1}\cdots u_{nx}^{l_n}~:~ \sum_{j=0}^n(j+2)l_j=k,~l_j\in {\mathbb Z}_{\ge 0}~\right\}.
\]
and ${\mathcal Q}_{k}[u]$ consists of the polynomials of weight $k$ containing the integral operator $D^{-1}$,
\[
{\mathcal Q}_{2n+5}[u]\subset {\rm Span}_{\mathbb R}\left\{X_{i_0}^{(n_0)}D^{-1}
X_{i_1}^{(n_1)}\cdots D^{-1}X_{i_l}^{(n_l)}:
\begin{array}{ll}
&\sum_{j=0}^l n_j+2l=n \\
&X_{i_j}^{(n_j)}\in {\mathcal P}_{2n_j+5}[u]
\end{array}
\right\}.
\]
The space ${\mathcal P}_k[u]$ can be also extended for $k=$even, and
 we will later define ${\mathcal Q}_k[u]$ for $k=$even.

\begin{Remark}
In order to verify the expansion (\ref{eq:pkdv}), one may need to impose
the following conditions for the initial data on $u(x,t=0)$,
\[ \left\{
\begin{array}{lll}
&{\rm a)}~|u(x,0)|\le C\exp(-\epsilon^{1/2}|x|), \quad {\rm as}\quad |x|\to \infty, \\
&{\rm b)}~\Vert u(x,0)\Vert^2_{H^{\infty}({\mathbb R})}=\displaystyle{\sum_{n=0}^{\infty}
\int_{\mathbb R}|u_{nx}(x,0)|^2 dx < \infty}.
\end{array}\right.
\]
The regorous justification of the KdV equation from a physical model such as the shallow 
water waves has been discussed in \cite{craig:85, kano:86}.
\end{Remark}

\section{Conserved quantities and $N$ soliton solutions}

The integrability of the KdV equation implies the existence of an infinite number of conserved quantities or integrals. Here we briefly summarize those quantities, and discuss the approximate integrals for the perturbed equation (\ref{eq:pkdv}). The approximate integrals will play a fundamental role for the
nomal form theory discussed in Section \ref{Snormalform}, and provide
an analyical tool to study the nonitegrable effect on the solution of the perturbed KdV equation.

\subsection{Conserved quantities}
Let us first recall the definition of a conserved quantity for 
the evolution equation in the form,
\begin{equation}
u_{t}=f(u),\quad {\rm with} \quad f(u)\in{\widehat{\mathcal P}}_{{\rm odd}}[u].\label{eq:conabst}
\end{equation}
\begin{Definition}
An integral of a differential polynomial $\rho(u)\in{\mathcal P}[u]=\oplus_k {\mathcal P}_k[u]$, 
\[
I[u]=\int_{{\mathbb R}}\rho(u)~dx,\quad {\rm so~that}\quad \rho(u) \in {\mathcal P}[u]/{\rm Im}(D),
\]
is a conserved quantity of (\ref{eq:conabst}) if 
\[
\rho_{t} \in {\rm Im}(D), \quad {\rm with} \quad D=\partial /\partial x.\]
The polynomial $\rho(u)$ is called a conserved density for (\ref{eq:conabst}). 
\end{Definition}
We also define a vector field generated by $f(u)$ and its action on the space ${\mathcal P}[u]$. 
\begin{Definition}
A vector field generated by $f(u)$ is defined by
\[
V_{f}:=\sum^{\infty}_{i=0}D^{i}(f)\frac{\partial}{\partial u_{ix}},
\]
which acts on the space ${\mathcal P}[u]$ as a differential operator,
\[\begin{array}{llll}
V_f: &{\mathcal P}[u] &\longrightarrow &{\mathcal P}[u] \\
&g &\longmapsto &V_f\cdot g
\end{array}
\]
\end{Definition}
With this definition, the condition for $\rho(u)$ being a conserved quantity can be expressed by 
\[
\rho_{t}=V_{f}\cdot \rho \in {\rm Im}(D),
\]
The KdV equation with $f(u)=K^{(0)}(u) \in {\mathcal P}_{5}[u]$ has an infinite sequence of conserved quantities,
\[ I_k^{(0)}[u]=\int_{\mathbb R}\rho_k^{(0)}(u)~dx, \quad {\rm with}\quad \rho_k^{(0)}\in{\mathcal P}_{2k+2}[u]\quad{\rm for}~~
k=0,1,2,\cdots.\]
which are generated by the bi-hamiltonian relation,
\begin{equation}
\label{bihamiltonian}
D\nabla I^{(0)}_{l+1}(u)=\Theta \nabla I^{(0)}_{l}(u),\quad
{\rm with}\quad \Theta=D^3+2(Du+uD).
\end{equation}
The gradient $(\nabla I)(u)$ is defined by
\[
\int_{\mathbb R}v (\nabla I) (u) ~dx =\lim_{\delta\to 0}\frac{d}{d\delta}I[u+\delta v],
\]
which can be expressed as
\[
(\nabla I)(u) =\sum_{i=0}^{\infty}(-1)^iD^i{\partial \rho \over \partial u_{ix}}(u), \quad {\rm for}\quad I[u]=\int_{\mathbb R}\rho(u)~dx.\]
The first few conserved densities $\rho_k^{(0)}$ are given by
\[\left\{\begin{array}{ll}
& \displaystyle{\rho^{(0)}_{0}=\frac{1}{2}u ,\quad 
\rho^{(0)}_{1}=\frac{1}{2}u^2}, \\
& \displaystyle{\rho^{(0)}_{2}=\frac{1}{2}(u_{x}^2-2 u^3), \quad 
\rho^{(0)}_{3}=\frac{1}{2}(u_{2x}^2-10 u u_{x}^2+5 u^4).} 
\end{array}\right.\]
Each density $\rho_k^{(0)}(u)$ can be considered as
\[
\rho_k^{(0)}\in {\mathcal P}_{2k+2}[u]/{\rm Im}(D).
\]
We then define
\begin{Definition}
The set of differential polynomials for the conserved densities are given by
\[ {\mathcal F}_{k}[u]\cong {\mathcal P}_{k}[u]/{\rm Im}(D), \]
where ${\mathcal F}_{k}[u]$ are defined by
\[
{\mathcal F}_{k}[u]:={\rm Span}_{\mathbb R}\left\{u^{l_{0}}u^{l_{1}}_{x}u^{l_{2}}_{2x} \cdots u^{l_{n}}_{nx}~: 
~\sum_{j=0}^n(j+2)=k,~l_{n}\ge 2\right\}.
\]
\end{Definition}
The following table shows the relation between the weight and the
dimension of the space ${\mathcal F}_k[u]$:\\

\medskip
\begin{center}
Table~1: The relation between the weight and the dimension of ${\mathcal F}_k[u]$
\end{center}
\begin{center}
 \begin{tabular}{|c|c|c|c|c|c|c|c|c|c|c|c|c|c|c|c|c|} \hline
Weight & 4 & 5 & 6 & 7 & 8 & 9 & 10 & 11 & 12 & 13 & 14 & 15 & 16 & 17 &
  18 & 19 \\ \hline
Dimension & 1 & 0 & 2 & 0 & 3 & 1 & 4 & 2 & 7 & 3 & 10 & 7 & 14 & 11 &
22  & 17 \\ \hline
  \end{tabular}
\end{center}
\medskip

\subsection{Approximate conserved quantities}
Here we discuss the conserved quantities of the perturbed KdV
equation $u_t+K(u;\epsilon)=O(\epsilon^{N+1})$ in (\ref{eq:pkdv}). We look for the conserved quantity in a formal power series, 
\[
I_{l}[u;\epsilon]=\int_{\mathbb R}\rho_{l}(u;\epsilon)~dx =I_l^{(0)}[u]+\epsilon I^{(1)}_{l}[u]+\cdots +\epsilon^N I_l^{(N)}(u)+O(\epsilon^{N+1}),
\]  so that the density $\rho_l(u;\epsilon)$ satisfies
\begin{equation}
 V_{K}\cdot \rho_{l}(u;\epsilon) = O(\epsilon^{N+1})\quad {\rm mod}~{\rm Im}(D). \label{eq:defofi}
\end{equation}
Here the higher order corrections are given by
\[
I^{(i)}_{l}[u]=\int_{\bf R}\rho^{(i)}_{l}(u) ~dx,\quad
\rho^{(i)}_{l}\in {\mathcal F}_{2(l+i)+2}[u]. 
\]
Expanding (\ref{eq:defofi}) in the power of $\epsilon$ leads to the equation for $\rho^{(m)}_{l}$,
\begin{equation}
V_{K_0^{(0)}}\cdot \rho^{(m)}_{l}(u)=-\sum_{i=1}^{m}V_{K^{(i)}}\cdot 
\rho^{(m-i)}_{l}(u), \quad {\rm for}\quad m=1,2,\cdots. 
\label{eq:recursioni}
\end{equation} 
Then we have,
\begin{Lemma}
\label{kerV}
For the linear map $V_{K^{(0)}_0}$,
\[
V_{K^{(0)}_0}~:~{\mathcal F}_{2k}[u]~~\longrightarrow~~{\mathcal F}_{2k+3}[u],
\]
the kernel of $V_{K^{(0)}_0}$ with a fixed weight is a one dimensional subspace of ${\mathcal F}_{2k}[u]$ given by
\[
{\rm Ker}V_{K^{(0)}_0}\cap {\mathcal F}_{2k}[u] = {\rm Span}_{\mathbb R}\{ \rho^{(0)}_{k-1}\}.
\]
\end{Lemma}
A proof of this Lemma can be found in \cite{novikov:84}. Namely there is only one conserved density in the form of differential polynomial for each weight.
From this lemma, we obtain a sufficient condition for the
solvability of $\rho^{(m)}_{l}$ on the space of differential polynomials ${\mathcal F}_{2(m+l)+2}[u]$, 
\begin{equation}
{\rm dim}~{\mathcal F}_{2(m+l)+5}[u]={\rm dim}~{\mathcal F}_{2(m+l)+2}[u]-1.
\end{equation}
The equation (\ref{eq:defofi}) is overdetermined in general, and
we denote by $N_l^{(m)}$ the number of the constraints for the existence of $\rho^{(m)}_{l}$, that is, 
\[
N_l^{(m)}={\rm dim}~{\mathcal F}_{2(m+l)+5}[u]-({\rm dim}~{\mathcal F}_{2(m+l)+2}[u]-1).
\] Note here that $N_{l_1}^{(m_1)}=N_{l_2}^{(m_2)}$ if $l_1+m_1=l_2+m_2$.
Then from the Table 1,
we obtain 
\begin{Proposition}
\label{PRO:conservation}
For the existence of the higher order corrections of the conserved quantities, we have
\begin{itemize}
\item[i)] There always exist conserved quantities, $I_{0}, I_{1}, I_{2}, I_{3}$ up to order $\epsilon$.\label{i}
\item[ii)] At order $\epsilon^2$, there exist $I_{0}, I_{1}, I_{2}$,
and $N_3^{(2)}=1$, that is, there is one condition $\mu_1^{(2)}=0$
   for the existence of $I_{3}$ where
\[\begin{array}{lll}
   \mu^{(2)}_{1} :=& -560a^{(2)}_{1}+170 a^{(2)}_{2}-60 a^{(2)}_{3}-8
    a^{(2)}_{4}+24 a^{(2)}_5-9 a^{(2)}_{6}+3a^{(2)}_{7}+24 a^{(2)}_{8} \\
&+{10\over 3} a^{(1)}_{1}(240 a^{(1)}_{1}-67a^{(1)}_{2}+6a^{(1)}_{4})+
{1\over 3} a^{(1)}_{2}(4a^{(1)}_{2}+30 a^{(1)}_{3}+a^{(1)}_{4})\\
&-a^{(1)}_{3}(2 a^{(1)}_{3}+a^{1}_{4}),
\end{array}
\]
\item[iii)] At order $\epsilon^3$, there always exist $I_{0}, I_{1} $. There are total three conditions, $\mu_k^{(3)}=0,~ k=1,2,3,$ with $\mu^{(3)}_{1}=0$ for the existence of {$I_{2}$} (i.e. $N_2^{(3)}=1$), and two
  conditions $\mu^{(3)}_{2}=0,~\mu^{(3)}_{3}=0$ for the existence of { $I_{3}$} (i.e. $N_3^{(3)}=2$). (Explicit form of $\mu_k^{(3)},~k=1,2,3,$ are listed in Appendix.)
\item[iv)] At order $\epsilon^4$, there always exist {$I_{0}$}. There are total seven conditions, $\mu_k^{(4)}=0~k=1,\cdots,7$ with $N_1^{(4)}+N_2^{(4)}+N_3^{(4)}=1+2+4=7$.
\end{itemize}
\end{Proposition}
One should note here that many physical examples have several conserved quantities, such as a total mass, momentum and energy,
which may be assigned as the first three quantities $I_l[u;\epsilon]$
with $l=0,1$ and $2$. Then the existence of the higher conservated
quantity $I_3[u;\epsilon]$ is a key for the integrability of the
perturbed equation. The item i) in Proposition \ref{PRO:conservation} suggests the (asymptotic) integrability of the perturbed equation (\ref{eq:pkdv}) up to order $\epsilon$.
In fact, in Section \ref{Snormalform}, we transform the perturbed equation to an integrable system up to order $\epsilon$, and discuss the effect of the nonexistence of $I_3^{(2)}$, i.e. $\mu^{(2)}_1\ne 0$, on the interaction behavior of two solitons.

\subsection{$N$ soliton solution}
One of the most important aspects of the existence of conserved quantities in the form of differential polynomial is to provide several exact solutions of the system. For example, $N$-soliton solution can be obtained by the variational equation (see for example \cite{novikov:84}),
\begin{equation}
\label{variationaleq}
{\nabla}{\mathfrak S}\{\lambda_1,\cdots,\lambda_N\}=0,
\end{equation}
 where ${\mathfrak S}\{\lambda_1,\cdots,\lambda_N\}$ is the invariants given by
\[{\mathfrak S}=I_{N+1}^{(0)}[u]+\lambda_1 I_{N}^{(0)}[u]+\cdots +\lambda_{N} I_1^{(0)}[u].\]
Here $\lambda_k$ are real constants (the Lagrange multipliers). 
Thus the $N$-soliton solution is given by a stationary point of the surface defined by $I_{N+1}^{(0)}[u]=$constant subject to the constraints, $I_k^{(0)}[u]=$constant for $k=1,2,\cdots,N$. This characterization of the $N$-soliton solution is essencial for its Lyapunov stability.

The equation (\ref{variationaleq}) gives a $2N$ order differential equation for $u(x,\cdot)$, and contains one soliton solution in the form,
\begin{equation}
\label{onesoliton}
u(x,\cdot)=2\kappa^2{\rm sech}^2(\kappa(x-x_0(\cdot))),
\end{equation}
with a appropriate choice of the parameters $\lambda_k$.
In fact, using the bi-hamiltonian relation (\ref{bihamiltonian})
the variational equation $\nabla {\mathfrak S}=0$ can be written as
\[({\mathfrak R}^{N-1}+\mu_1{\mathfrak R}^{N-2}+\cdots+\mu_{N-1})
D(\nabla I_2^{(0)}+\mu_N\nabla I_1^{(0)})=0,\]
where $\mu_k$'s are some constants related to $\lambda_j$.
Then by choosing $\mu_N=-4\kappa^2$ this equation
 admits the one soliton solution in the form (\ref{onesoliton}).  Here the operator ${\mathfrak R}$ is called the recursion operator defined as
\begin{equation}
\label{recursionop}
 {\mathfrak R}=\Theta D^{-1}, \quad {\rm and}\quad {\mathfrak R} D\nabla I_k^{(0)}=D\nabla I_{k+1}^{(0)}.
\end{equation}
The recursion operator plays an important role for the theory of the integrable systems.

\section{Symmetry and the perturbed equation}
Here we review the symmetries of the KdV equation and
discuss how these symmetries work for the analysis of the nearly
integrable systems.

\subsection{Symmetries of the KdV equation}
Let us define the symmetry for a system,
\begin{equation}
u_{t}=K(u), \quad {\rm with}\quad K(u)\in \widehat{\mathcal P}[u].
\label{eq:evolution}
\end{equation}
\begin{Definition}
\label{symmetry}
A function $S(u)\in \widehat{\mathcal P}[u]$ is a symmetry of (\ref{eq:evolution}) if $S(u)$ satisfies 
the commutation relation 
\[
{\rm ad}_K\cdot S(u):=[K,S](u)= (V_S\cdot K- V_K\cdot S)(u)=0, \label{eq:symmetrycondition}
\]
where ${\rm ad}_K:\widehat{\mathcal P}[u]\to \widehat{\mathcal P}[u]$ is the adjoint map of $K$. 
\end{Definition}
This means that if $S(u)$ is a symmetry of $K(u)$ then the flows of 
vector fields generated by $K(u)$ and $S(u)$ commute with each other,
i.e. $V_K\circ V_S-V_S\circ V_K=0$. 

The KdV equation has an infinite number of commuting symmetries which are given by the hamiltonian flows generated by the conserved quantities $I_l^{(0)}[u]$ for $l=1,2,\cdots$, that is,
\[ K_0^{(n)}(u):=D\nabla I_{n+2}^{(0)}, \quad {\rm for}\quad n=-1,0,1,2,\cdots.\]
Then using the recursion operator ${\mathfrak R}$ in (\ref{recursionop}) those symmetries can be constructed as,
\[K_0^{(n-1)}(u)={\mathfrak R}^n K_0^{(-1)}(u), \quad{\rm with}\quad K_0^{(-1)}(u)=u_x.\]

The symmetry can be also constructed by the so-called {\it master symmetry} \cite{master}. 
The definition of the master symmetries is as follows. 
A function $M(u)$ is a master symmetry of the evolution equation 
(\ref{eq:evolution}) if the Lie bracket defined by 
$M$ maps symmetries onto symmetries, i.e.,
\[
[M,S_{i}]=S_{j}, 
\]
where $S_{i}$ and $S_{j}$ are the symmetries.

The explicit form of the first few commutative symmetries $K_{0}^{(i)},~i \ge -1,$ are
\[\begin{array}{llll}
K_0^{(-1)}&=&u_{x},  \\
K_0^{(0)}&=&6uu_{x}+u_{3x}, \quad {\rm the~KdV~equation}\\
K_0^{(1)}&=&u_{5x}+10u_{3x}u+20u_{2x}u_{x}+30u_{x}u^2,  \\
K_0^{(2)}&=&u_{7x}+14u_{5x}u+42u_{4x}u_{x}+70u_{3x}u^2+70u_{3x}u_{2x}+280u_{2x}u_{x}u\\
&{}&+140u_{x}u^3+70u_{x}^3, \\
K_0^{(3)}&=&u_{9x}+18u_{7x}u+72u_{6x}u_{x}+168u_{5x}u_{2x}+126u_{5x}u^2+252u_{4x}u_{3x}\\
&{}&+756u_{4x}u_{x}u+1260u_{3x}u_{2x}u+966u_{3x}u_{x}^2+420u_{3x}u^3\\
&{}&+1302u_{2x}^2u_{x}+2520u_{2x}u_{x}u^2+1260u_{x}^3u+630u_{x}u^4, \\
\end{array}\]
and the master symmetries $M_{0}^{(j)},~j \ge 0,$ are
\[\begin{array}{lll}
M_0^{(0)}&=&xK_0^{(-1)}+2u, \\
M_0^{(1)}&=&xK_0^{(0)}+8u^2+4u_{2x}+2K_0^{(-1)}D^{-1}(u),  \\
M_0^{(2)}&=&xK_0^{(1)}+32u^3+48uu_{2x}+36u_{x}^2+6u_{4x}+2K_0^{(0)}D^{-1}(u)\\
&{}&  +6K_0^{(-1)}D^{-1}(u^2), \\
M_0^{(3)}&=&xK^{(2)}_0+8u_{6x}+96uu_{4x}+240u_{x}u_{3x}+160u_{2x}^2+384u^2u_{2x}+576uu_{x}^2\\
&{}&+128u^4+2K_0^{(1)}D^{-1}(u)+6K_0^{(0)}D^{-1}(u^2)+10K_0^{(-1)}D^{-1}(2u^3-u_{x}^2).
\end{array}\]

The set of those polynomials $\{K^{(n)}_0,M^{(m)}_0\}$ forms an infinite dimensional Lie algebra containing a classical Virasoro algebra of the master symmetries, that is,
\begin{equation}
\label{virasoro}
  [M_{n},K_{m}]=(2m+3)K_{n+m}, \quad 
  [M_{n},M_{m}]=2(m-n)M_{n+m}.
\end{equation}
One note here that the symmetries $K_0^{(n)}(u)$ are the odd weight (differential) polynomials, while the master symmetries are the even weight polynomials. In particular, each $M_0^{(k)}(u)-xK_0^{(k-1)}(u)$ can be considered as an element of the space of even weight polynomials generated by the recursion formula starting from $k=0$,
\[\begin{array}{lll}
Y^{(k)}(u)&=&\alpha_1^{(k)}u_{2kx}  \\
&{}&+\displaystyle{\sum_{k_1+k_2=k-1 \atop { 1\le i\le N(k_1) \atop
1\le j\le N(k_2)}}\left(\beta^{(k)}_{ij}(Y_i^{(k_1)}Y_j^{(k_2)})(u)+
\gamma^{(k)}_{ij}(X_i^{(k_1-1)}D^{-1}Y_j^{(k_2)})(u)\right)},
\end{array}\]
where $X_i^{(k)}(u)$ is a monomial in $\widehat{\mathcal P}_{2k+5}[u]$.
This defines the space $\widehat{\mathcal P}_{{\rm even}}[u]$,
\[
\widehat{\mathcal P}_{{\rm even}}[u]=\bigoplus_{k=0}^{\infty}
{\widehat{\mathcal P}}^{(k)}_{{\rm even}}[u], \quad {\rm with}\quad
\widehat{\mathcal P}^{(k)}_{{\rm even}}[u]={\mathcal P}^{(k)}_{{\rm even}}[u]\oplus
{\mathcal Q}^{(k)}_{{\rm even}}[u],\]
where ${\mathcal P}^{(k)}_{{\rm even}}[u]={\mathcal P}_{2k+2}[u]$.
The space $\widehat{\mathcal P}_{{\rm even}}^{(k)}[u]$ will be important for the normal form theory discussed in the next section.
\begin{Remark}
The actions of the symmetries and the master symmetries on one-soliton solution
(\ref{onesoliton}) give a representation of the algebra (\ref{virasoro})
in terms of the soliton parameters $\kappa$ and $\theta=\kappa x_0$:
For the symmetry $K_0^{(n)}$, the action generates the vector field,
\[ V_n=\displaystyle{4^{n+1}\kappa^{2n+3}{\partial \over \partial \theta},\quad {\rm for}\quad n\ge -1,}\]
and for the master symmetry,
\[ W_m=\displaystyle{4^n\kappa^{2n+1}{\partial \over\partial \kappa}, \quad {\rm for}\quad m\ge 0.}\]
Those gives a representation of the algebra (\ref{virasoro}), that is,
\[
[V_n,V_m]=0,\quad [W_n,V_m]=(2m+3)V_{n+m}, \quad [W_n,W_m]=2(n-m)W_{n+m}.\]
\end{Remark}

\subsection{Approximate symmetries}
As we discussed on the approximate conserved quantity for the perturbed equation (\ref{eq:pkdv}) in the previous section, one can also discuss the approximate symmetries for (\ref{eq:pkdv}).
We say that a differential polynomial $H(u)\in {\mathcal P}[u]$ is an approximate symmetry of order $n$ of the perturbed equation (\ref{eq:pkdv}), if the commutator of $K(u)$ and $H(u)$ gives
\[ [K,H](u)=O(\epsilon^{n+1}).\]
Expanding $H(u)$ in the power series of $\epsilon$, 
$H=H^{(0)}+\epsilon H^{(1)}+\epsilon^2 H^{(2)}+\cdots$,
we have the equation for $H^{(m)}(u)$,
\[  {\rm ad}_{K^{(0)}_0}\cdot H^{(m)} =-\sum_{j=1}^m[K^{(j)},H^{(m-j)}], \quad {\rm for}\quad m=1,2,\cdots,\]
where $K^{(0)}_0=u_{3x}+6uu_x$, and $K^{(j)}$'s are the higher order corrections of the KdV equation (\ref{eq:pkdv}). Then choosing $H^{(0)}$ be one of the symmetries of the KdV equation, say $H^{(0)}=K_0^{(n)}$,
we find the obstacles for the existence of higher order corrections $H^{(m)}$ for the approximate symmetry, and obtain the same conditions for the existence as stated in Proposition \ref{PRO:conservation} \cite{mikhailov:91}.
In the proof, one needs the following Lemma similar to Lemma \ref{kerV} for the kernel of the adjoint action ${\rm ad}_{K^{(0)}_0}$,
\begin{Lemma}
\label{keradK}
The kernel of ${\rm ad}_{K^{(0)}_0}$ on the space of differential polynomials ${\mathcal P}_{2n+5}[u]$ is given by
\[
{\rm Ker}\left({\rm ad}_{K^{(0)}_0}\right)\cap{\mathcal P}_{2n+5}[u]={\rm Span}_{\mathbb R}\{K^{(n)}\}, \quad {\rm for}\quad n\ge -1.\]
\end{Lemma}

\section{Normal form theory}
\label{Snormalform}
The normal form theory has been well developed in the study of finite dimensional dynamical systems
(cf.\cite{arnold}). The main purpose of the normal form is to classify 
the vector fields near critical points in terms of the symmetries
of the leading order equation. This concept has been applied for the perturbed
KdV equation as well as the nonlinear Schr\"odinger equation in
\cite{kodama:85, bbmnormal, kano:89, obstacle}.

In this section we review the normal form theory for the perturbed
KdV equation \cite{bbmnormal, kdvnormal}, and discuss the effect of the obstacles on two-soliton
interactions. We also discuss the Gardner-Miura transformation (which is an invertible version of the Miura transformation) in terms of the normal form theory, and show that the Gardner-Miura transformation is just a normal form transformation.

\subsection{Normal form}
The basic idea of the normal form for the perturbed KdV equation is to
remove all the nonresonant (nonsecular) terms in the higher order corrections using a near identity transformation given by the Lie transformation. 
Since the symmetries $K_0^{(n)}(u)$ of the KdV equation give the obvious resonant terms, we write each higher order term $K^{(n)}(u)$ in the form,
\[
K^{(n)}(u)=a_1^{(n)}K_0^{(n)}(u)+F^{(n)}(u), \quad {\rm for}\quad n=1,2,\cdots,
\]
so that $F^{(n)}(u)$ has no linear term. Then the point of the transformation
is to simplify the term by removing the nonresonant terms in $F^{(n)}(u)$. If one succeeded to remove the entire $F^{(n)}(u)$ up to $n=N$, then the perturbed
equation is {\it asymptotically} integrable up to the order $\epsilon^N$, and it possesses approximate integrals $I_l^{(n)}[v;\epsilon]$ for all $l\in{\mathbb Z}_{\ge 0}$ and $n=1,\cdots,N$. However as we have shown in 
Proposition \ref{PRO:conservation}, there are conditions for the existence
of approximate conserved quantities. Thus we expect to see {\it obstacles}
in removing all the nonlinear terms $F^{(n)}(u)$ from the higher order corrections. In fact we have
\begin{Lemma}
\label{lietransform}
There exists a near identity transformation, $T_{\epsilon}:v\mapsto u$,
\[ u=T_{\epsilon}(v)=u+\epsilon \phi^{(1)}(u) +\cdots, \quad {\rm with}
\quad \phi^{(n)}(u)\in\widehat{\mathcal P}^{(n)}_{{\rm even}}[u]\]
such that the perturbed equation (\ref{eq:pkdv}) is transformed to
\begin{eqnarray}
&&v_{t}+G(v;\epsilon)=O(\epsilon^{N+1}), \label{eq:gkdv}\\
&&\quad {\rm with}\quad G(u;\epsilon)=K^{(0)}_0(v)+\epsilon G^{(1)}(v)+\epsilon^2 G^{(2)}(v)+\cdots + \epsilon^{N} G^{(N)}(v), 
\nonumber
\end{eqnarray} 
where $G^{(n)}(u)$ are given by
\[
G^{(n)}(v)=a_1^{(n)}K_0^{(n)}(v) + R^{(n)}(v), \quad {\rm with}\quad
R^{(n)}(u)=\sum_{i=1}^{\Delta(n)}\mu_i^{(n)}R^{(n)}_i(u).\]
Here the constants $\mu_i^{(n)}$ are given in Proposition \ref{PRO:conservation} for the existence of the approximate conserved quantities, $\Delta(n)$ is the total number of the conditions for the existence, and some $R_i^{(n)}(u)\in \widehat{\mathcal P}_{2n+5}[u]$.
\end{Lemma}

\begin{Proof}
We take the Lie (exponential) transform for a near identity transform, that is,
\[
u=T_{\epsilon}(v)=\exp V_{\phi} \cdot v, \
\]
where the generating function $\phi$ is expanded in the power series of $\epsilon$,
\[
\phi=\epsilon \phi^{(1)} + \epsilon^2 \phi^{(2)} + \cdots +\epsilon^N \phi^{(N)} +O(\epsilon^{N+1}).
\]
Substituting this into (\ref{eq:pkdv}) leads to 
\[
V_{G}={\rm Ad}_{\exp{V_{\phi}}} \cdot V_{K} := \exp V_{\phi} \cdot V_{K} \cdot \exp (-V_{\phi}), \]
which gives 
\[
\displaystyle{G=\sum_{n=0}^{\infty}{1\over n!}\left({\rm ad}_{\phi}\right)^n\cdot K=K + 
[\phi,K]+\frac{1}{2!}[\phi,[\phi,K]]+\frac{1}{3!}[\phi,[\phi,[\phi,K]]]+ \cdots.}\]
Then from each order of $\epsilon$, we obtain
\[ 
\begin{array}{llll}
&[K^{(0)}_0,\phi^{(1)}] &=& K^{(1)}-G^{(1)}, \\ 
&[ K_0^{(0)}, \phi^{(2)}] &=& K^{(2)}-G^{(2)}+\frac{1}{2}[\phi^{(1)},K^{(1)}+G^{(1)}], \\ 
&[ K_0^{(0)}, \phi^{(3)}] &=& K^{(3)}-G^{(3)}+\frac{1}{2}[\phi^{(2)},K^{(1)}+G^{(1)}], \\ 
&&{}&+\frac{1}{2}[\phi^{(1)},K^{(2)}+G^{(2)}]+\frac{1}{12}[\phi^{(1)},[\phi^{(1)},K^{(1)}+G^{(1)}]], \\ 
& &\cdot & \cdots .
\end{array} \]
Thus we have the equation for $\phi^{(n)}$ in the form called the {\it homological} equation,
\begin{equation}
\label{homologicaleq}
{\rm ad}_{K_0^{(0)}}\cdot \phi^{(n)}:=[K_0^{(0)},\phi^{(n)}]=\tilde K^{(n)}-G^{(n)},
\end{equation}
where $\tilde K^{(n)}$ is successively determined from the previous equations.
 Since the ad-action ${\rm ad}_{K^{(0)}_0}$ raises the weight by three, we have
\[
{\rm ad}_{K^{(0)}_0}: \widehat{\mathcal P}_{2n+2}[v] 
\longrightarrow \widehat{\mathcal P}_{2n+5}[v]. \]
Then we assume
\[ \begin{array}{lll}
\phi^{(1)}&=&\alpha^{(1)}_{1}u^2+\alpha^{(1)}_{2}u_{2x}+\alpha^{(1)}_{3}u_{x}D^{-1}(u), \\
&{}\\
\phi^{(2)}&=&\alpha^{(2)}_{1}u^3+\alpha^{(2)}_{2}uu_{2x}+\alpha^{(2)}_{3}u^2_{x}+
\alpha^{(2)}_{4}u_{4x}\\
&{}& +\alpha^{(2)}_{5}K^{(0)}_0 D^{-1}(u)+\alpha^{(2)}_{6}u_{x}D^{-1}(u^2), \\
&{}\\
\phi^{(3)}&=&\alpha^{(3)}_{1}u^{4x}+\alpha^{(3)}_{2}uu^{2}_{x}+\alpha^{(3)}_{3}u^{2}u_{2x}+
\alpha^{(3)}_{4}u^{2}_{2x}+\alpha^{(3)}_{5}u_{x}u_{3x}\\
&{}&+\alpha^{(3)}_{6}uu_{4x}+\alpha^{(3)}_{7}u_{6x}+\alpha^{(3)}_{8}u_{x}D^{-1}(u^{3})
+\alpha^{(3)}_{9}u_{x}D^{-1}(u^{2}_{x}) \\
&{}& + \alpha^{(3)}_{10}K^{(0)}_0D^{-1}(u^{2})+\alpha^{(3)}_{11}K_{0}^{(1)}D^{-1}(u)+\alpha^{(3)}_{12}K^{(1)}D^{-1}(u).
\end{array}\]
The homological equation (\ref{homologicaleq}) gives a linear system of equations for the column vector $\alpha^{(n)}:=(\alpha^{(n)}_1,\cdots,\alpha^{(n)}_{N(n)})^T$ with $N(n)={\rm dim}~\widehat{\mathcal P}_{2n+2}[v]$, 
\begin{equation}
\label{matrixrep}
 A\alpha^{(n)}=b^{(n)},
 \end{equation}
where $A$ is a $(M(n)-1)\times N(n)$ matrix representation of the linear map ${\rm ad}_{K^{(0)}_0}$ on $\widehat{\mathcal P}_{2n+2}[v]$,
and $b^{(n)}$ represents the coefficients of the monomial in $\tilde K^{(n)}-G^{(n)}$  which has $M(n)-1$ elements with $M(n)={\rm dim}~\widehat{\mathcal P}_{2n+5}[v]$.
The system (\ref{matrixrep}) is overdetermined, and the total number of constraints for the consistency of the system is given by 
\[
N{(n)}:={\rm dim}~\widehat{\mathcal P}_{2n+5}-1 -{\rm dim}~\widehat{\mathcal P}_{2n+2}.\]
This number should agree with that of the conditions for the existence of 
approximate symmetries, that is, the number of $\mu^{(n)}_i$ in 
Proposition \ref{PRO:conservation}.

Let us give an explicit form of $G^{(n)}$ up to $n=2$:

At order $\epsilon$, the matrix $A$ in (\ref{matrixrep}) is given by a $3\times 3$ matrix with rank $3$, so that we have $R^{(1)}=0$, that is, no obstacle. The explicit transformation $\phi^{(1)}(v)$ is given by
\[\left\{\begin{array}{lll}
&\displaystyle{\alpha^{(1)}_{1}=\frac{1}{6}(20a^{(1)}_{1}+a^{(1)}_{2}-a^{(1)}_{4}),
\quad\alpha^{(1)}_{2}=\frac{1}{12}(10a^{(1)}_{1}+a^{(1)}_{3}-a^{(1)}_{4}),}\\
&\displaystyle{\alpha^{(1)}_{3}=\frac{1}{3}(10a^{(1)}_{1}-a^{(1)}_{2}).}
\end{array}\right.
\]

At order $\epsilon^2$, the matrix $A$ is an $7 \times 6$ matrix with rank $6$, and we have one obstacle with the form
\[\begin{array}{lll}
 R^{(2)}(v)&=& b^{(2)}_{2}v_{5x}v+b^{(2)}_{3}v_{4x}v_{x}+
b^{(2)}_{4}v_{3x}v^{2}+b^{(2)}_{5}v_{3x}v_{2x} \\
&{}& +~b^{(2)}_{6}v_{2x}v_{x}v+
b^{(2)}_{7}v_{x}v^{3}+b^{(2)}_{8}v^{3}_{x}, \end{array} \]
Here $b^{(2)}_{i},~i=2,3,\cdots,8$, are constants satisfying the condition,
\[
170b^{(2)}_{2}-60b^{(2)}_{3}-8b^{(2)}_{4}+24b^{(2)}_{5}-9b^{(2)}_{6}
+3b^{(2)}_{7}+24b^{(2)}_{8}=1. \]
The explicit formula of the $\alpha^{(2)}=(\alpha_1^{(2)},\cdots,\alpha_6^{(2)})^T$ 
is given in Appendix.
\end{Proof}

The perturbed equation (\ref{eq:pkdv}) may have several approximate conserved quantities based on the original physical setting. Then we consider a particular form of the transformed equation (\ref{eq:gkdv}) whose conserved quantities are given by those of the KdV equation.
We call this form of equation the normal form of (\ref{eq:pkdv}), that is, we define
\begin{Definition}
\label{kdvnormalform} 
For a subset of integers $\Gamma\subset {\mathbb Z}_{\ge 0}$, suppose that the perturbed KdV equation (\ref{eq:pkdv}) has the approximate
conserved quantities $I_{l}[u;\epsilon]$ for $l \in \Gamma$ up to order $\epsilon^N$.
Then the normal form of (\ref{eq:pkdv}) is defined by (\ref{eq:gkdv}),
\[
v_t + \sum^{N}_{n=0} \epsilon^n G^{(n)}(v)=O(\epsilon^{N+1}), \]
whose conserved
quantities $J_{l}[v;\epsilon]:=I_l[T_{\epsilon}(v);\epsilon]$ for $l \in \Gamma$ are expressed in terms of the conserved quantities of the KdV equation $I^{(0)}_{l}[v]$,
\begin{equation}
\label{jconservation}
J_{l}[v;\epsilon]=I^{(0)}_{l}[v]+\epsilon c_{l}^{(1)}I^{(0)}_{l+1}[v]+
\cdots+\epsilon^N c_l^{(N)}I^{(0)}_{N+l}[v] 
+O(\epsilon^{N+1}), 
\end{equation}
 where $c_l^{(i)},~i=1,2,\cdots,N$, are some real constants.
\end{Definition}
In particular, if the set $\Gamma$ contains the first three numbers,
$\Gamma\supseteq\{0,1,2\}$, then the normal form admits a solitary wave solution in the form of KdV soliton (\ref{onesoliton}).
This can be seen by taking the variation,
\[\nabla (J_2[v;\epsilon]+\lambda J_1[v;\epsilon])=0.\]
Since many physically interesting systems possess those conserved
quantities as mass, momentum and energy, we expect to find a solitary wave close to the KdV soliton for such systems.
Now we show the existence of such normal form for the case $\Gamma=\{0,1,2\}$:
\begin{Proposition}
\label{explicitform}
Suppose that the perturbed KdV equation (\ref{eq:pkdv}) has the first three approximate conserved quantities, $I_l[u;\epsilon],~l=0,1,2$
up to order $\epsilon^3$. Then the corresponding normal form takes the form (\ref{eq:gkdv}) with 
\[\begin{array}{lll}
R^{(1)}(v)&=& 0, \\
R^{(2)}(v)&=& \mu_1^{(2)}R^{(2)}_1, \\
R^{(3)}(v)&=& \mu_1^{(2)} c_2^{(1)}{\mathfrak R}(R^{(2)}_1)
+\mu_2^{(3)}R_1^{(3)}+\mu_3^{(3)}R_2^{(3)},
\end{array}\]
where the conserved quantities $J_l[v,\epsilon]$ for the normal form are given by
\[
J_l[v;\epsilon]=I_l[v]+\epsilon c_l^{(1)} I_{l+1}[v] +\cdots +\epsilon^3 c_l^{(3)} I_{l+3}[v]+O(\epsilon^4).\]
The obstacles $R^{(n)}_k$ are expressed as
\[ \begin{array}{llll}
R^{(2)}_{1}&=&\frac{-1}{50}\left(v_{5x}v+\frac{3}{2}v_{4x}v_{x}+5v_{3x}v^2
-\frac{5}{2}v_{3x}v_{2x}+20v_{2x}v_{x}v+10v_{x}v^3+5v_{x}^3  \right),\\
&&{}\\
R^{(3)}_{1}&=&\frac{1}{175}\left(v_{7x}v+\frac{5}{2}v_{6x}v_{x}+14v_{5x}v^2-
\frac{7}{2}v_{4x}v_{3x}+63v_{4x}v_{x}v+35v_{3x}v_{2x}v\right. \\ &{}&+56v_{3x}v^{2}_{x}+42v_{3x}v^{3}+42v^{2}_{2x}v_{x}+273v_{2x}v_{x}v^2+168v^{3}_{x}v\\
&{}&\left.+\frac{105}{2}v_{x}v^{4}+21v_{x}D^{-1}(v^{3}_{x}) \right),  \\
&&{}\\
R^{(3)}_{2}&=&\frac{1}{175}\left(4v_{7x}v+10v_{6x}v_{x}+\frac{91}{2}v_{5x}v^{2}-14v_{4x}v_{3x}
+\frac{441}{2}v_{4x}v_{x}v \right.\\
&{}& \left.+\frac{385}{2}v_{3x}v_{2x}v +203v_{3x}v^{2}_{x}+147v_{3x}v^{3}+\frac{357}{2}v^{2}_{2x}v_{x}+903v_{2x}v_{x}v^2\right.\\
&{}&\left.+483v^{3}_{x}v+210v_{x}v^{4}+21v_{x}D^{-1}(v^{3}_{x})\right).
\end{array}
\]
and ${\mathfrak R}$ is the recursion operator.
\end{Proposition}
\begin{Proof}
Recall that $J_l[v,\epsilon]=\int_{\mathbb R}\rho(v;\epsilon)dx$ is an approximate conserved quantity for (\ref{eq:gkdv}) if
\[
V_G\cdot \rho(v;\epsilon)+O(\epsilon^{N+1})\in {\rm Im}(D).\]
Then using the form (\ref{eq:gkdv}) and $\rho_l=\rho_l^{(0)}+\epsilon c_l^{(1)}\rho_{l+1}^{(0)}+\cdots$, we have, at order $\epsilon^2$,
\begin{equation}
\label{obstacle2}
V_{R^{(2)}}\cdot \rho_l^{(0)}(v)\in {\rm Im}(D), \quad {\rm for}\quad l=0,1,2,
\end{equation}
and at order $\epsilon^3$,
\begin{equation}
\label{obstacle3}
\left\{ \begin{array}{ll}
{}&V_{R^{(3)}} \cdot \rho_k^{(0)}(v)\in {\rm Im}(D), \quad {\rm for} \quad k=0,1, \\
{}&V_{R^{(3)}}\cdot \rho_2^{(0)}+c_2^{(1)}V_{R^{(2)}}\cdot \rho_3^{(0)}\in {\rm Im}(D).
\end{array}\right.
\end{equation}
Then from a direct computation with the explicit form of 
$R^{(2)}\in {\mathcal P}_{9}[v]$, the conditions (\ref{obstacle2}) 
lead to the required form of $R^{(2)}(v)$.

For (\ref{obstacle3}), we first write $R^{(3)}$ in the sum of homogeneous solution $R_h^{(3)}$ and a particular solution $R_p^{(3)}$, that is,
\[ \left\{
\begin{array}{ll}
& V_{R_h^{(3)}}\cdot \rho_l^{(0)}\in {\rm Im}(D), \quad {\rm for}\quad l=0,1,2\\
&V_{R_p^{(3)}}\cdot \rho_2^{(0)}+c^{(1)}_2V_{R^{(2)}}\cdot\rho_3^{(0)}\in {\rm Im}(D).
\end{array}\right.\]
Then one can find a particular solution by using the recursion operator ${\mathfrak R}=\Theta D^{-1}$,
\[\begin{array}{llll}
V_{R^{(2)}}\cdot \rho_3^{(0)}&=& R^{(2)}\nabla I_3^{(0)} & {\rm mod}~{\rm Im}(D) \\
&=& R^{(2)}D^{-1}D\nabla I_3^{(0)} & {\rm mod}~{\rm Im}(D) \\
&=& R^{(2)}D^{-1}\Theta \nabla I_2^{(0)} & {\rm mod}~{\rm Im}(D) \\
&=& (\Theta D^{-1}R^{(2)})\nabla I_2^{(0)} & {\rm mod}~{\rm Im}(D) \\
&=& V_{{\mathfrak R}R^{(2)}}\cdot \rho_2^{(0)} & {\rm mod}~{\rm Im}(D) \end{array}\]
from which we have
\[
R^{(3)}_p=-\mu_1^{(2)}c_2^{(1)}{\mathfrak R}(R^{(2)}_1).
\]
Then from a direct computation we have the homogeneous solution in the desired form,
\[
R^{(3)}_h=\mu_2^{(3)}R^{(3)}_1+ \mu_3^{(3)}R^{(3)}_2.\]
\end{Proof}

\begin{Remark}
We have the following remarks on the obstacles:
\begin{itemize}
\item[a)]
In general, the explicit form of the obstacles $R^{(n)}$ may be successively obtained by solving the linear equations,
\[
V_{R^{(n)}}\cdot \rho_{l}^{(0)}+\sum_{k=1}^{n-2}c_l^{(k)}V_{R^{(n-k)}}\rho_{l+k}^{(0)}\in {\rm Im}(D), \quad {\rm for}\quad n=2,3,\cdots.\]
Suppose we found $R^{(k)}$ up to $k=n-1$. Then we have a particular solution in the form,
\[R^{(n)}_p=-\sum_{k=1}^{n-2}c_l^{(k)}{\mathfrak R}^k R^{(n-k)}. \] 
Here one has to check the compatibility among the different $l=0,1,2$.
\item[b)]
Since the normal form with $\Gamma=\{0,1,2\}$ admits the solitary wave in the form of the KdV soliton, one can see that all the obstacles vanish when $v$ assumes the KdV soliton solution,
\[
R^{(n)}(v)=0, \quad {\rm when}\quad v=2\kappa^2 {\rm sech}^2(\kappa(x-x_0)).\]
\end{itemize}
\end{Remark}

\subsection{The Gardner-Miura transformation}
It was found in \cite{miura:68} that there is an {\it invertible} transformation 
between the KdV equation and the KdV equation with a cubic nonlinear term,
\begin{equation}
\label{gardnereq}
u_{t}+6uu_{x}+u_{3x}=\epsilon au^2u_{x}, 
\end{equation}
where $a$ is an arbitrary constant.
The transformation is called the Gardner-Miura transformation which is an invertible 
version of the Miura transformation,
\begin{equation}
\label{gardnertrans}
v=u - \alpha \epsilon^{1/2} u_{x} - \alpha^2 \epsilon u^2,\quad
{\rm with}\quad \alpha = -{\sqrt{a/6}} .
\end{equation}
Here we treat (\ref{gardnereq}) as an example of the perturbed KdV equation, and give an explicit formulation of the Gardner-Miura transformation in terms of the normal form theory.
Since the perturbed equation (\ref{gardnereq}) has an infinite number of conserved densities and there is no resonant term as the symmetry of the KdV equation in the higher order, the normal form is just the KdV equation. Then we construct the normal form transformation,
$u=T_{\epsilon}(v)$ which is the inverse of the Gardner-Miura transformation 
(\ref{gardnertrans}).

 Since the Gardner-Miura transformation (\ref{gardnertrans}) is of a Riccati type, the change of the variable
\begin{equation}
 u = \frac{1}{\delta}\left(D\ln \varphi+\frac{1}{2 \delta} \right), \quad{\rm with}\quad \delta=\alpha \epsilon^{1/2}, \label{eq:dlnphiu}
\end{equation}
leads to the Schr{\"o}dinger equation,
\[
L^2 \varphi:=(D^2+v)\varphi=k^2\varphi, \quad {\rm with}\quad k=-1/(2\delta).\label{schrodinger}
\]
Using the notion of the pseudo-differential operators $D^{\nu}$ for $\nu\in{\mathbb Z}$, one can define $L=(D^2+v)^{1/2}$ as
\[
L=D+q_{1}D^{-1}+q_{2}D^{-2}+\cdots,\quad q_{i}\in {\mathcal P}_{i+2}[v] , \]
Then writing $D$ in the power series of $L$, we have
\[
D=L+p_{1}L^{-1}+p_{2}L^{-2}+\cdots,~~~~~~~p_{i}\in {\mathcal P}_{i+2}[v] \]
and using $L\varphi=k\varphi$, we find
\begin{equation}
\label{delphi}
D \ln \varphi=-\frac{1}{2\delta}+\sum^{\infty}_{i=1}(-2\delta)^{i}p_{i}(v). 
\end{equation}
Note in particular that we have
\[ q_1={1\over 2}v, \quad p_1=-{1\over 2}v.\]
From (\ref{eq:dlnphiu}), the equation (\ref{delphi}) leads to the inverse of the Gardner-Miura transformation,
\[\begin{array}{lll}
u&=&\displaystyle{\sum^{\infty}_{i=1}(-2)^i\delta^{i-1}p_{i}(v,v_{x},\cdot\cdot\cdot) } \\
&=&v+\delta v_{x}+\delta^2(v_{2x}-v^2)+ \cdot \cdot \cdot. 
\end{array}
\]
We now remove the terms of the non-integer powers of
$\epsilon$, the odd integer of $\delta=\alpha \epsilon^{1/2}$, in this equation.
The first term  $\delta v_{x}$ can be removed by 
the translation of $x$. After removing the term $\delta v_x$, we have
$K_0^{(0)}(v)$ at the order of $\epsilon^{3/2}$. 
This can be removed by the translation of $t$. Continuing this process,
we see the symmetries of the KdV equation at the non-integer powers of $\epsilon$. Then shifting the symmetry parameters $x_{2n+1}$, those can be removed. Here $x=x_1$ and $t=-x_3$. Now we can show
\begin{Proposition}
\label{gardnernormaltrans}
The inverse of the Gardner-Miura transformation gives a normal form transform,
\[
u=\frac{1}{\delta}D^{-1}\sinh \left(\delta \sum^{\infty}_{i=0}\frac{\delta^{2i}}{2i+1}
 \ \frac{\partial}{\partial x_{2i+1}} \right)v, \quad \delta=-\sqrt{{\epsilon a\over 6}},
\]
where the derivative of $v$ with respect to $x_{2n+1}$ defines the symmetry, that is, $\partial v/ \partial x_{2n+1}=K_0^{(n-1)}(v)$.
\end{Proposition}
\begin{Proof}
It is well known \cite{satotheory} that the wave function $\varphi$ of 
the Schr{\"o}dinger equation can be expressed by the $\tau$-function, 
\[
\varphi({\bf x},k)=\frac{\tau\left(
{\bf x}+2\langle \delta \rangle \right)}{\tau({\bf x})} \ {\rm e}^{kx_1}, \quad {\rm with}\quad k=-{1\over 2\delta},
\]
where we denote ${\bf x}:=(x_1, x_2, \cdots)$, and 
\[
\langle \delta\rangle =\left(\delta,{\delta^3\over 3},{\delta^5\over 5},\cdots\right).\]
Expanding the equation $D\ln \varphi$ with this equation in the power of $\delta$, we find 
\begin{equation}
\label{vtau}
 v({\bf x})=2D^2 \ln \tau({\bf x}) .
\end{equation}
Then from (\ref{eq:dlnphiu}) we have
\[
u({\bf x})=\frac{1}{\delta}D\left(\ln \tau({\bf x}+2\langle \delta \rangle)- \ln \tau({\bf x}) \right).
\]
Now applying the vertex operator
\[
V(\delta)=\exp\left(-\delta \sum^{\infty}_{i=0}\frac{\delta^{2i}}{(2i+1)} \ 
\frac{\partial}{\partial x_{2i+1}}  \right), 
\]
and using (\ref{vtau}), we obtain the result.
Note here that both $u({\bf x})$ and $u({\bf x}+\langle \delta\rangle)$ satisfy the same equation (\ref{gardnereq}).
\end{Proof}

\section{Interactions of solitary waves}
As an important application of the normal form theory, we consider the
interaction problem of
solitary waves, and show how the theory enables us to understand the
interaction properties
under the influence of the higher orders in the perturbed KdV equation
(\ref{eq:pkdv}).
We assume that the perturbed equation possesses the first three conserved
quantities, i.e.
$\Gamma=\{0,1,2\}$. Then the first obstacle appears in the order
$\epsilon^2$ as $\mu_1^{(2)}\ne 0$, that is, we consider the normal form,
\[
v_t+K_0^{(0)}(v)+\epsilon a_1^{(1)}K_0^{(1)}(v) +\epsilon^2
(a_1^{(2)}K_0^{(2)}(v)+
\mu_1^{(2)}R^{(2)}_1(v))=0.
\]
Several physical examples of this type will be discussed in Section
\ref{chap:examples}.
We use the method of perturbed inverse scattering
transform \cite{pismkaup, pism} to analyze the normal form.
We start with a brief description of the method.

\subsection{Inverse scattering transform}
The key of the inverse scattering transform is based on the one-to-one
correspondence for each $t$ between the scattering data $S(t)$ and the
potential $v(x,t)$, the solution of the normal form, of the
Schr\"odinger equation (see for example \cite{ist}),
\[
\frac{\partial^2 \varphi}{\partial x^2}+(v+k^2)\varphi=0.
\]
The correspondence is given by the formula
of $v(x,t)$ in terms of the squared eigenfunctions,
\begin{equation}
\label{vphi2}
v(x,t)=4\sum_{j=1}^N\kappa_j(t)C_j(t)\varphi^2(x,t;i\kappa_j)+\frac{2i}{\pi}
\int_{\mathbb R}kr(t;k)\varphi^2(x,t;k)~dk.
\end{equation}
Here the scattering data $S(t)$ is defined by
\[
S(t):=\left\{\{\kappa_j(t)> 0,C_j(t)\}_{j=1}^N,~ r(t;k)~{\rm for}~
k\in{\mathbb R}\right\}, \]
and the eigenfunction $\varphi(x,t;k)$ is assumed to satisfy the boundary
condition,
\[
\varphi(x,t;k) \longrightarrow e^{-i k x},~~~~~{\rm as}~~x \longrightarrow
-\infty . \]
Note that the squared function $\varphi^2(x,t;k)$ satisfies
\[
\Theta \varphi^2=(D^3+2(Dv+vD))\varphi^2=-4k^2D\varphi^2,\]
so that the function $D\varphi^2$ is the eigenfunction of the recursion
operator ${\mathfrak R}$ with the eigenvalue $-4k^2$.

With the formula of $v$ in (\ref{vphi2}), the conserved quantities
$J_i[v;\epsilon]$ can be expressed in terms of the scattering data,
\begin{equation}
\label{integrals}
J_{i}[v;\epsilon]=I^{(0)}_{i}[v]+\epsilon c^{(1)}_{i} I^{(0)}_{i+1}[v]
+\epsilon^2 c^{(2)}_{i} I^{(0)}_{i+2}[v]+O(\epsilon^3),\quad i=0,1,2,
\end{equation}
where $I^{(0)}_{j}$ can be expressed as
\begin{equation}
\label{kdvintegral}
I^{(0)}_{m}=\frac{2^{2m+1}}{2m+1}(-1)^{m+1}
\left[\sum^{M}_{j=1}\kappa^{2m+1}_{j}-(-1)^{m}\Delta_{m}\right]
\end{equation}
with the radiation part $\Delta_m(t)$,
\[
\Delta_{m}=-\frac{2m+1}{2\pi}\int^{\infty}_{0}k^{2m}\ln (1-|r(t;k)|^2)~dk \
\ge 0.
\]
The existence of such conserved quantities plays a crucial rule for the
interaction mechanism as in the case of the KdV solitons.

The time evolution of the scattering data is determined by
\[
\frac{d}{dt}S(t)=-V_K\cdot S(t), \quad {\rm for}\quad v_t+K(v)=0 .
\]
In particular, the equation for the reflection coefficient $r(t;k)$ is given by
\[
\frac{d}{dt}r(t;k)=\frac{-1}{2ika^2(t;k)}\int_{{\mathbb
R}}v_{t}(x,t)\phi^2(x,t;k)~dx.
\]
where $a(t;k)$ is the reciprocal transmission coefficient determined by
$r(t;k)$ and $\kappa_j$.
Using the normal form $v_t+K(v)=0$, we obtain the equation for $r(t;k)$,
\[
\frac{d r}{dt}=i \omega r + \epsilon^2 \frac{\mu^{(2)}_{1}}{2 i k
a^2}\int_{\mathbb R} R^{(2)}_{1} \varphi^2(x,k)~ dx +
O(\epsilon^3),\label{eq:rhoevo}
\]
where $\omega=8k^3(1-4\epsilon a^{(1)}_{1}k^2+16a^{(2)}_{1}\epsilon^2k^4)$.

Those given above provide an enough information for our purpose of studying
the interaction of solitary waves.

\subsection{solitary wave interaction}
 Recall that the obstacle vanishes for one soliton solution of the KdV
equation, that is, there is no effect of the obstacle on the solitary wave.
The higher order
 terms lead to the shift of the velocity of the soliton solution due to the
resonance caused by the symmetries of the KdV equation \cite{kodamataniuti}. In
order to see the effect of obstacles, we now consider the interaction of two
solitary waves, and show the inelasticity in the interaction which can be
considered as a nonintegrable effect of the obstacle.

\subsubsection{Inelasticity in the interaction}
Let us assume that the initial data consists of two well-separated solitary
waves with parameters $\kappa_1 >\kappa_2$ in the form of
(\ref{onesoliton}), traveling with speed $s_j=4\kappa_j^2(1-4\epsilon
a_1^{(1)}\kappa_j^2+16\epsilon^2 a_1^{(2)}\kappa_j^4)~j=1,2$, and
approaching each other. Then we analyze the interaction by using a
perturbation method where the leading order solution is assumed to be the
exact two-soliton solution of the KdV equation, that is, for large $x_{01}-x_{02}\gg 1$
\[
v(x,0)\simeq  2\kappa_1^2 {\rm sech}^2(\kappa_1 (x-x_{01})) +
2\kappa_2^2 {\rm sech}^2(\kappa_2(x-x_{02})).
\]
We then determine the evolution of the scattering data, $\kappa_j(t)$ and
the radiations $\Delta_m(t)$ in (\ref{eq:deltam}).
First we have
\begin{Proposition} \cite{bbmnormal}
\label{kappashift}
Due to the interaction of two solitary waves, their parameters
$\kappa_j~j=1,2$ are shifted by $\Delta\kappa_j(t)$ which can be expressed
in terms of the radiations $\Delta_m(t)$,
\[
\Delta \kappa_{1}=\frac{5\kappa_{2}^2 \Delta_{1}+3 \Delta_{2}}{15
\kappa_{1}^2(\kappa_{1}^2-\kappa_{2}^2)}+O(\epsilon\Delta_m) \ge 0, \quad
\Delta \kappa_{2}=-\frac{5\kappa_{1}^2 \Delta_{1}+3 \Delta_{2}}{15
\kappa_{2}^2(\kappa_{1}^2-\kappa_{2}^2)}+O(\epsilon\Delta_m) \le 0. \]
There is also a production of the third soliton with the parameter,
\[
\Delta\kappa_{3}=\frac{15 \kappa_{1}^2 \kappa_{2}^2 \Delta_{0}+5
(\kappa_{1}^2+\kappa_{2}^2)\Delta_{1}+3\Delta_{2}}{15\kappa_{1}^2
\kappa_{2}^2}+ O(\epsilon\Delta_m) \ge 0.
\]
\end{Proposition}
\begin{Proof}
 From the conserved quantities $J_l[v;\epsilon]=$constant for $l=0,1,2$ up
to order $\epsilon^2$, we obtain,
\[
\Delta I_l^{(0)}+\epsilon c_l^{(1)}\Delta I_{l+1}^{(0)} +\epsilon^2
c_l^{(2)}\Delta I_{l+2}^{(0)}=O(\epsilon^3),
\]
where the variations $\Delta I_n^{(0)}$ are taken over the shifts $\Delta
\kappa_j$ and the radiation $\Delta_m$, that is,
\[
(2l+1)\sum_{j=1}^N\kappa_j^{2l}\Delta\kappa_j-(-1)^l\Delta_l=O(\epsilon
\Delta_n),
\quad {\rm for}\quad l=0,1,2.
\]
Here $N-2$ is a possible number of new solitary waves. Since the
$\Delta\kappa_j$
for $j>2$ represents a new eigenvalue of the Schr\"odinger equation, it is
nondegenerate and an isolated point on the imaginary axis of the spectral
domain . This implies that the number of new eigenvalues should be just one
for a sufficiently small $\epsilon$, i.e. $N=3$.

Because of $\kappa_j=0$ initially for $j>2$,  one can find the formulae of
$\Delta\kappa_j$ for $j=1,2$ from those variations for $l=1,2$. Also the
first variation with $l=0$,
\[
\sum_{j=1}^3 \Delta\kappa_j-\Delta_0=O(\epsilon\Delta_n),
\]
leads to the formula $\Delta\kappa_3$ for a new solitary wave.
\end{Proof}
Thus we find the followings:
\begin{itemize}
\item[a)] The total mass, $M=\int vdx\propto \kappa$, of the larger solitary
wave is increased, and contrary that of the smaller solitary wave is
decreased by the interaction.
\item[b)] The amount of the energy change, $E=\int v^2dx\propto\kappa^3$,
has the property
$1<|\Delta E_2/\Delta E_1|<(\kappa_1/\kappa_2)^2$, i.e. the energy expense
of the smaller solitary wave is more than the energy gain of the larger one.
\item[c)] The interaction produces a new solitary wave as well as radiation.
\end{itemize}
Those results except for a new solitary wave production are consistent with
the numerical
observation in \cite{bbmnumerical}. It may be difficult to observe the new
solitary wave from the numerical calculation, since this solitary wave has
long width and small amplitude $(\Delta\kappa_3)^2$ which is of order
$\epsilon^8$ (see below).

The function $\Delta_m(t)$ of the radiation can be computed as follows:
The reflection coefficient $r(t;k)$ in (\ref{eq:rhoevo}) can be expressed as
\[
r(t;k)=\frac{\epsilon^2 \mu^{(2)}_{1}}{2 i k}\left( \int^{t}_{0}d\tau \frac
{{\rm e}^{-i \omega \tau}}{a(\tau;k)^2} \int_{{\mathbb
R}}dx\varphi^2(x,\tau;k)R^{(2)}_{1}(v) \right){\rm e}^{i \omega t}.
\]
Then $\Delta_{m}(t)$ is given by
\begin{equation}
\label{eq:deltam}
\Delta_{m}(t)=\frac{2m+1}{2\pi}\int^{\infty}_{0}D_{m}(t;k)~dk+o(\epsilon^4),
\end{equation}
where $D_{m}(t;k)$ is
\[\begin{array}{lll}
D_{m}(t;k)&=&k^{2m}|r(t;k)|^2 + o(\epsilon^4) \\
&=&\displaystyle{\epsilon^4\frac{(\mu^{(2)}_{1})^2 k^{2(m-1)}}{4}\left|
\int^{t}_{0}d\tau~
{\rm e}^{-i \omega \tau}\int_{{\mathbb R}}dx
~\varphi^2(x,\tau;k)R^{(2)}_{1}(v) \right|^2
+o(\epsilon^4).}
\end{array}\]
We numerically calculate the formula $\Delta_m(t)$ by means of perturbation,
that is,
we assume $v(x,t)$ to be
a two-soliton solution of the integrable one
$v_{t}+K^{(0)}+\epsilon a^{(1)}_{1}K_{1}+\epsilon^2 a^{(2)}_{1}K_{2}
=0$ and $\varphi(x,t,k)$ is the corresponding eigenfunction. The result will
be shown in Section 7 for the examples of ion acoustic waves and the Boussinesq equation.

\subsubsection{The additional phase shifts of solitary wave}
\label{phaseshift}
As a consequence of the nonlocal terms in the normal form transformation, one can
find the additional phase shifts on the solitary waves $u(x,t)$ of the
perturbed KdV equation (\ref{eq:pkdv}) through their interaction.

Let us first recall the phase shifts of the two-soliton solution for
$v(x,t)$ \cite{gardner:74, ablowitz:82}. The asymptotic form of $v(x,t)$
consists of well separated one solitons,
\[\begin{array}{lll}
v(x,t) &\approx&\displaystyle{  v_j^{\pm}(x,t), \quad {\rm as} \quad\left\{
{t\to \pm\infty}\atop{\kappa_j x\sim\omega_jt}\right.} \\
&{}& {\rm with}\quad v_j^{\pm}(x,t)=2\kappa^2{\rm
sech}(\kappa_jx-\omega_jt-\theta_j^{\pm}).
\end{array}\]
Then the phase shifts $\Delta x_j^{(0)}:=(\theta_j^+-\theta_j^-)/\kappa_j$
are given by
\[
\Delta x_1^{(0)}=\displaystyle{-{1\over \kappa_1}\ln
\left({\kappa_1-\kappa_2\over\kappa_1+\kappa_2},\right)}
\quad {\rm and}\quad
\Delta x_2^{(0)}=\displaystyle{{1\over \kappa_2}\ln
\left({\kappa_1-\kappa_2\over\kappa_1+\kappa_2}\right)}
\]
We now compute the correction to the shift $\Delta x_j^{(0)}$ using the
normal form transformation.
  From the asymptotic form of the two-soliton solution for $t\to -\infty$, we
have
up to order $\epsilon$
\[\begin{array}{llll}
u(x,t)&\underset{\kappa_1x\sim\omega_1 t}{\longrightarrow}
&\displaystyle{ v_1^-+\epsilon \left(\alpha_1^{(1)}(v_1^-)^2+\alpha_2^{(1)}
(v_1^-)_{2x}+\alpha_3^{(1)}(v_1^-)_x\int_{-\infty}^xv_1^-~dx\right)}, \\
& \underset{\kappa_2x\sim\omega_2 t}{\longrightarrow}&
\displaystyle{v_2^-+\epsilon \left(\alpha_1^{(1)}(v_2^-)^2+\alpha_2^{(1)}
(v_2^-)_{2x}+\alpha_3^{(1)}(v_2^-)_x\int_{-\infty}^xv_2^-~dx\right)}\\
&{}&\displaystyle{\quad +\epsilon\alpha_3(v_2^-)_x\int_{\mathbb R}v_1^-~dx.}
\end{array}\]
One should note here that there is an extra term in the $v_2$ solitary wave
which is the key term for the additional phase shift. Namely the term can be
absorbed as a translation of $x$ in $v_2$. The other terms contribute to
modify the shape of the soliton, the dressing part.

Also for $t\to +\infty$, we have
\[\begin{array}{llll}
u(x,t)&\underset{\kappa_1x\sim\omega_1 t}{\longrightarrow}
&\displaystyle{ v_1^++\epsilon \left(\alpha_1^{(1)}(v_1^+)^2+\alpha_2^{(1)}
(v_1^+)_{2x}+\alpha_3^{(1)}(v_1^+)_x\int_{-\infty}^xv_1^+~dx\right)}\\
&{}&\displaystyle{\quad +\epsilon\alpha_3(v_1^+)_x\int_{\mathbb R}v_2^+~dx.}\\
& \underset{\kappa_2x\sim\omega_2 t}{\longrightarrow}&
\displaystyle{v_2^++\epsilon \left(\alpha_1^{(1)}(v_2^+)^2+\alpha_2^{(1)}
(v_2^+)_{2x}+\alpha_3^{(1)}(v_2^+)_x\int_{-\infty}^xv_2^+~dx\right)}\\
\end{array}\]
Now the additional shift appears to $v_1$. This can be extended for the next
order where the shift also appears as a traslation of $t$
with a term like $K_0^{(0)}(v_1)\int_{\mathbb R}v_2dx$.
Thus we have the total phase shift for $v_1$ solitary wave,
\begin{equation}
\label{eq:phaseshift}
\Delta x_1=\Delta x_1^{(0)}+\epsilon \Delta x_1^{(1)}+\epsilon^2\Delta
x_1^{(2)}+O(\epsilon^3)
\end{equation}
where the additional phase shifts are given by
\[\begin{array}{lll}
\Delta x^{(1)}_{1}&=& \displaystyle{\alpha^{(1)}_{3}\int_{{\mathbb R}}v_{2}dx
=4\kappa_{2}\alpha^{(1)}_{3},} \\
\Delta x^{(2)}_{1}&=&\displaystyle{\alpha^{(2)}_{6}\int_{{\mathbb R}}(v_2)^2
dx+
\frac{\alpha^{(1)}_{3}(\alpha^{(1)}_{1}-\alpha^{(1)}_{3})}{2}\int_{{\mathbb R}}
(v_2)^2 dx + 4\kappa_{1}^2 \alpha^{(2)}_{5}\int_{{\mathbb R }}v_2 dx } \\
&=&\displaystyle{\left(\frac{16}{3}\alpha^{(2)}_{6}+
\frac{8}{3}\alpha^{(1)}_{3}(\alpha^{(1)}_{1}-\alpha^{(1)}_{3})\right)\kappa_
{2}^{3}+
16\alpha^{(2)}_{5}\kappa_{1}^2\kappa_{2}.}
\end{array}\]

\section{Examples}
\label{chap:examples}
 In this section, the normal form theory is applied to some
explicit models including the ion acoustic wave equation, the Bousinesq equation as a model of the shallow water waves, 
and the regularized long wave equation (sometimes called BBM equation). We also carry out the numerical simulation for those examples and compare the results with the predictions obtained from the normal form theory such as the phase shift (\ref{eq:phaseshift}) and the radiation energy (\ref{eq:deltam}). 

We also consider the 7th order Hirota KdV equation which is known to be nonintegrable even though it admits an exact two solitary wave solution. The main issue is to determine the order of the obstacle
of the corresponding normal form, which indicates an nonintegrability of the equation in the asymptotic sense. It turns out that the obstacles appear at order $\epsilon^4$.

\subsection{Ion acoustic waves}
An asymptotic property of the ion acoustic waves
has been discussed in several papers
(see for example \cite{taniuti}). These 
studies show that the KdV equation is 
derived as the first approximation of the ion acoustic wave equation 
under the weakly dispersive limit. The higher order corrections to the KdV soliton solution have been also discussed in \cite{kodamataniuti}.
Recently, Li and Sattinger in \cite{sattinger} studied numerically the interaction problem of solitary waves and showed 
that the amplitude of the radiation after two solitary wave interaction can be observed as small as $10^{-5}$ order, and they concluded that the KdV equation gives an excellent approximation.
Here we explain those observations based on 
the normal form theory developed in the previous sections.

Following the method in \cite{taniuti}, we first derive the KdV equation with the higher order corrections. The ion
acoustic wave equation is expressed by the system of
three partial differential equations in the dimesionless form, 
\begin{equation}
\label{eq:ionacoustic}
\left\{\begin{array}{lll}
&(n_{i})_{T}+(n_{i}v_{i})_{X}=0, \\
&\displaystyle{(v_{i})_{T}+\left( \frac{v^{2}_{i}}{2}+\phi \right)_{X}=0,} \\ 
&\phi_{2X}-\exp\phi+n_{i}=0 , 
\end{array}\right.
\end{equation}
where $n_{i}$, $v_{i}$, and $\phi$ are the normalized variables for ion density, ion velocity and  electric potential. Electron density $n_{e}$ is related with $\phi$ as $n_{e}=\exp(\phi)$. Assuming the weak nonlinearity and the weak dispersion, we introduce the scaled variables,
\[
v_i=\epsilon v, \quad n_e=1+\epsilon n, \quad {\rm and}\quad x=\epsilon^{1/2} X, \quad t=\epsilon^{1/2} T.\]
Then we write (\ref{eq:ionacoustic}) in the following form for 
$(n,v)$,
\[\left\{\begin{array}{lll}
&\displaystyle{\left(1-\epsilon D^2\cdot{1\over 1+\epsilon n}\right)n_t 
+v_x+\epsilon(nv)_x-\epsilon(v\ln (1+\epsilon n))_x=0,} \\
&{}\\
&\displaystyle{v_t+vv_x+{1\over 1+\epsilon n}n_x=0 .}
\end{array}\right. \]
Then inverting the operator in front of $n_t$, we have the matrix equation,
\begin{equation}
\label{eq:ionacousticbasic}
\frac{\partial}{\partial t}U + A_0\frac{\partial}{\partial x}U + \epsilon
\frac{\partial}{\partial x}B(U)=0, \quad {\rm with}\quad U=\left(
\begin{matrix} n \\ v \end{matrix}\right),
\end{equation}
where $A_0$ is the constant matrix $A_0= \left(\begin{matrix}
0 & 1\\ 1 &0 \end{matrix}\right)$, and the vector function $B(U)$ is given by the expansion
\[ \begin{array}{lll}
&B= B^{(1)}+\epsilon B^{(2)}+\epsilon^2 B^{(3)}+ \cdots, \\
&{}\\
&{\rm with} \left\{
\begin{array}{ll}
&B^{(1)}=\left(\begin{matrix}
v_{2x}+nv\\ {1\over 2}(v^2-n^2)\end{matrix}\right),\quad 
B^{(2)}=\left(\begin{matrix}
v_{4x}+n_xv_x \\ {1\over 3}n^3\end{matrix}\right), \\
&{}\\
&B^{(3)}=\left(\begin{matrix}
v_{6x}+(n_xv_x)_{2x}-(nv_{3x})_x-vnn_x \\ -{1\over 4}n^4 \end{matrix}\right) .
\end{array}\right.\end{array} \]
Thus for the case with $\epsilon=0$, we have two simple linear waves propagating with the
speeds $\lambda_{\pm}=\pm 1$ given by the eigenvalues of the matrix $A_0$. We then look for an asymptotic wave along with the speed 
$\lambda_+=1$, so that we introduce the scaled variables on this moving frame,
\[
x'=x-t, \quad t'=\epsilon t,\]
which gives, after dropping the prime on the new variable,
\begin{equation}
\label{eq:uequation}
\displaystyle{\left(A_0-I\right){\partial U \over \partial x}
+\epsilon\left({\partial U\over \partial t}+{\partial B(U)\over \partial x}\right)=0}.
\end{equation}
where $I$ is the $2\times 2$ identity matrix.
Let us decompose $U$ in the form,
\[
U(x,t)=u(x,t) R_{+}+ f(x,t) R_{-},\]
where $R_{\pm}$ are the eigenvectors corresponding to the eigenvalues
$\lambda_{\pm}=\pm 1$. Then taking the projections of (\ref{eq:uequation}) on the $R_{\pm}$-directions, we have
\[\left\{\begin{array}{lll}
&\displaystyle{ u_t+\left(L_+B\right)_x=0}, \\
&{} \\
&\displaystyle{-2f_x+\epsilon\left(f_t+(L_-B)_x\right)=0},
\end{array}\right.\]
where $L_{\pm}$ are the left eigenvectors with the normalization, $L_{\pm}R_{\pm}=1,~L_{\pm}R_{\mp}=0$.
Then we can see that $f(x,t)$ can be expressed in an expansion form, 
\[
f(x,t)=\epsilon f^{(1)}(u)+\epsilon^2 f^{(2)}(u)+\epsilon^3 f^{(3)}(u)+ \cdots, 
\quad {\rm with}\quad f^{(k)}(u)\in \widehat{\mathcal P}^{(k)}_{{\rm even}}[u].\]
where $f^{(n)}$ are determined iteratively from
\[
f=\displaystyle{{\epsilon\over 2}(L_-B)+{\epsilon\over 2}D^{-1}f_t=
{\epsilon\over 2}(L_-B^{(1)})+{\epsilon^2\over 2}(L_-B^{(2)}) +{\epsilon\over 2}f_t+\cdots}.
\]
Thus we obtain the perturbed KdV equation which takes the following form 
up to order $\epsilon^2$ after an appropriate normalization,
\[
\begin{array}{lll}
&&\displaystyle{u_{t}+6uu_{x}+u_{3x}+\epsilon\left(\frac{1}{4}u_{5x}-u_{3x}u-\frac{3}{2}u_{x}u^2\right) +\epsilon^2\left(\frac{5}{72}u_{7x}-\frac{1}{2}u_{5x}u \right.}\\
&& \quad\displaystyle{\left. -\frac{17}{24}u_{4x}u_{x}+\frac{1}{6}u_{3x}u_{2x}+\frac{3}{4}u_{3x}u^2-\frac{1}{4}u_{2x}u_{x}u -\frac{7}{8}u_{x}^3+\frac{1}{4}u_{x}u^3\right)=O}(\epsilon^3), 
\end{array}\]
 from which we have 
 \[
 \displaystyle{\mu^{(2)}_{1}=
-\frac{11}{9} \ne 0. }\]
 Thus the normal form of the ion acoustic wave equation has the obstacle  $R^{(2)}_{1}$. The normal form transformation is given by  
\[ \left\{
\begin{array}{lll}
&&\displaystyle{\alpha^{(1)}_{1}=\frac{11}{12},~~~~~\alpha^{(1)}_{2}=\frac{1}{3},~~~~~
\alpha^{(1)}_{3}=\frac{7}{6},}  \\
&&\\
&&\displaystyle{\alpha^{(2)}_{1}=\frac{731}{3600},~~~~~\alpha^{(2)}_{2}=\frac{89}{600},
~~~~~\alpha^{(2)}_{3}=\frac{433}{3600},} \\
&&\\
&&\displaystyle{\alpha^{(2)}_{4}=\frac{51}{800},~~~~~\alpha^{(2)}_{5}=\frac{23}{1800},
~~~~~\alpha^{(2)}_{6}=\frac{87}{400}.} 
\end{array} \right. \]
\begin{Remark}
The physical variables $n$ and $v$ are expressed as,
\[\displaystyle{n=u+f,\quad v=u-f, \quad{\rm with}\quad  R_{\pm}=\left(
\begin{matrix} 1 \\ {\pm}1\end{matrix}\right).}\]
Since $f$ has an expansion of the power series of $\epsilon$ and each coefficient $f^{(n)}$ is an element in $\widehat{\mathcal P}^{(n)}_{{\rm even}}$, the expressions of $n,~v$ have the same form as of the normal form transformation. In the asymptotic sense, all the physical variables are expressed by one function $u$ as $U=uR_+$. Then the choice of the higher order terms $f$ has a freedom. Then the main purpose of the normal form is to use this freedom to classify near integrable systems in the asymptotic sense.
\end{Remark}

We now compare the results in Section 6 with numerical results. 
We used the spectral method~\cite{spectral} for the numerical 
computation. For a convenience, we consider (\ref{eq:ionacoustic}) in a moving frame with a speed $c$,
\[\left\{
\begin{array}{lll}
&&(n)_{t}-(6+c)n_{x}+6v_{x}+6(n v)_{x}=0, \\
&&\displaystyle{(v)_{t}-(6+c)v_{x}+6\left( \frac{v^{2}}{2}+\phi \right)_{x}=0,} \\ 
&&\phi_{2x}-3\exp\phi+3(n+1)=0 , 
\end{array}\right.\]
The computation is done with the $2^{12}$ number of Fourier modes and the time step $dt=0.008$. In general, the spectral method yields eliasing errors 
from the high frequency modes in the nonlinear terms, so we try 
to get rid of their errors by adopting the $3/2$ rule~\cite{aliasing}.

Let us first comapare a solitary wave solution of the ion acoustic wave equation with the KdV soliton solution. 
Since the one-soliton solution (\ref{onesoliton}) is a kernel of the obstacle
$R^{(2)}_{1}$, one can construct a solitary wave for the variable $u$ by the normal form transformation, i.e.,
\[ \begin{array}{lll}
&\displaystyle{u=v+\epsilon \phi^{(1)}+\epsilon^2 \left(\phi^{(2)}+\frac{1}{2}V_{\phi^{(1)}}\cdot \phi^{(1)}\right) 
+ O(\epsilon^3), }\\
&{}\\
&\hspace*{1cm}{\rm with}\quad v=2\kappa^2{\rm sech}^2(\kappa(x-x_{0})), \label{eq:core}
\end{array}\]
where $x_{0}$ is a center position of the soliton at $t=0$.
The solitary wave for the variable $u$ is constructed by the core 
$v$ with the dressing terms $\epsilon \phi^{(1)}+\epsilon^2
(\phi^{(2)}+\frac{1}{2}V_{\phi^{(1)}}\cdot \phi^{(1)}) +
O(\epsilon^3)$. Now let us see the effect of the dressing terms, which is considered as nonresonant higher harmonics in a periodic solution of finite dimensional problem.
\begin{figure}[htbp]
\begin{center}
\vspace*{0.5cm}
 \epsfig{file=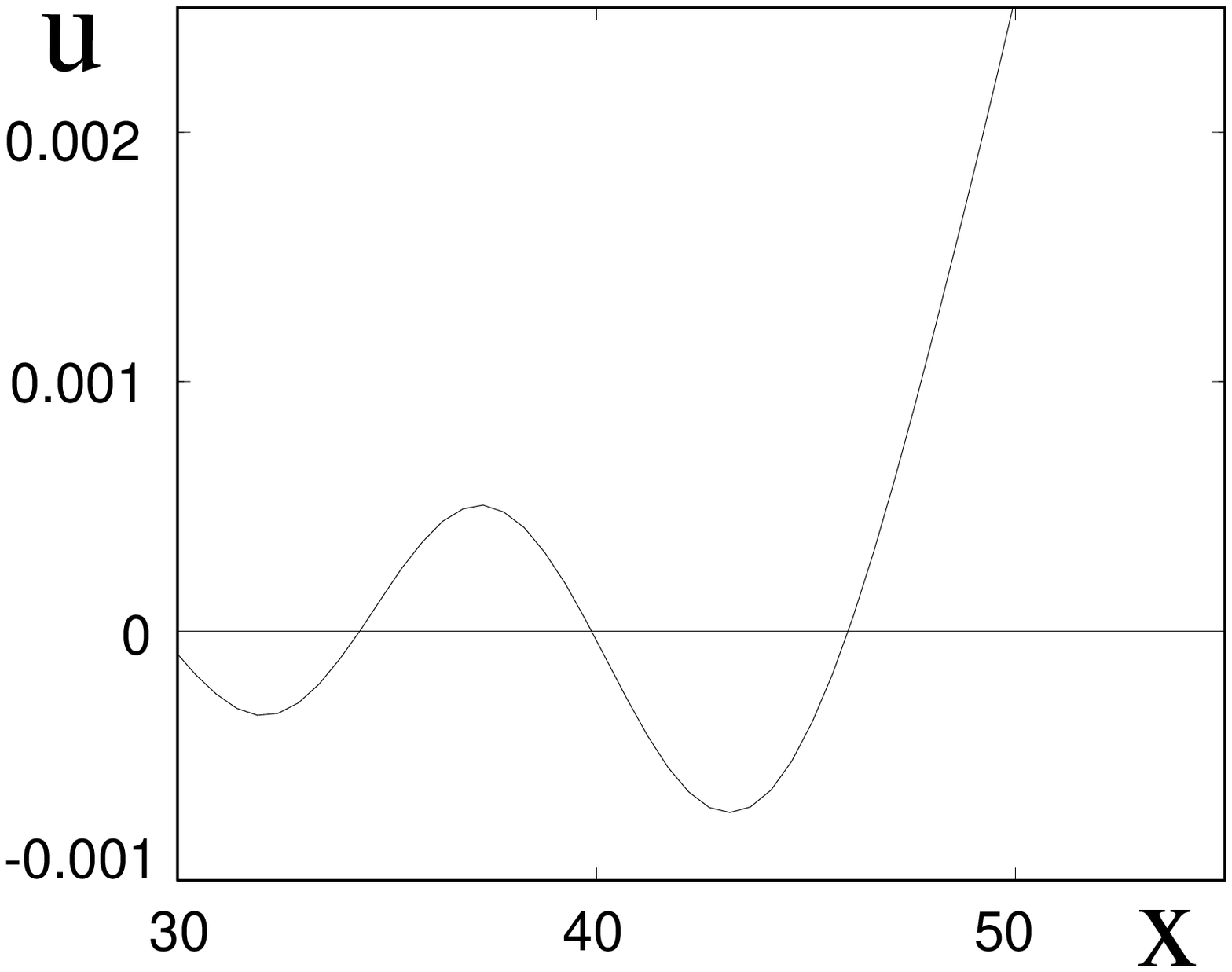,height=4cm,width=4cm}\hspace*{0.3cm}
 \epsfig{file=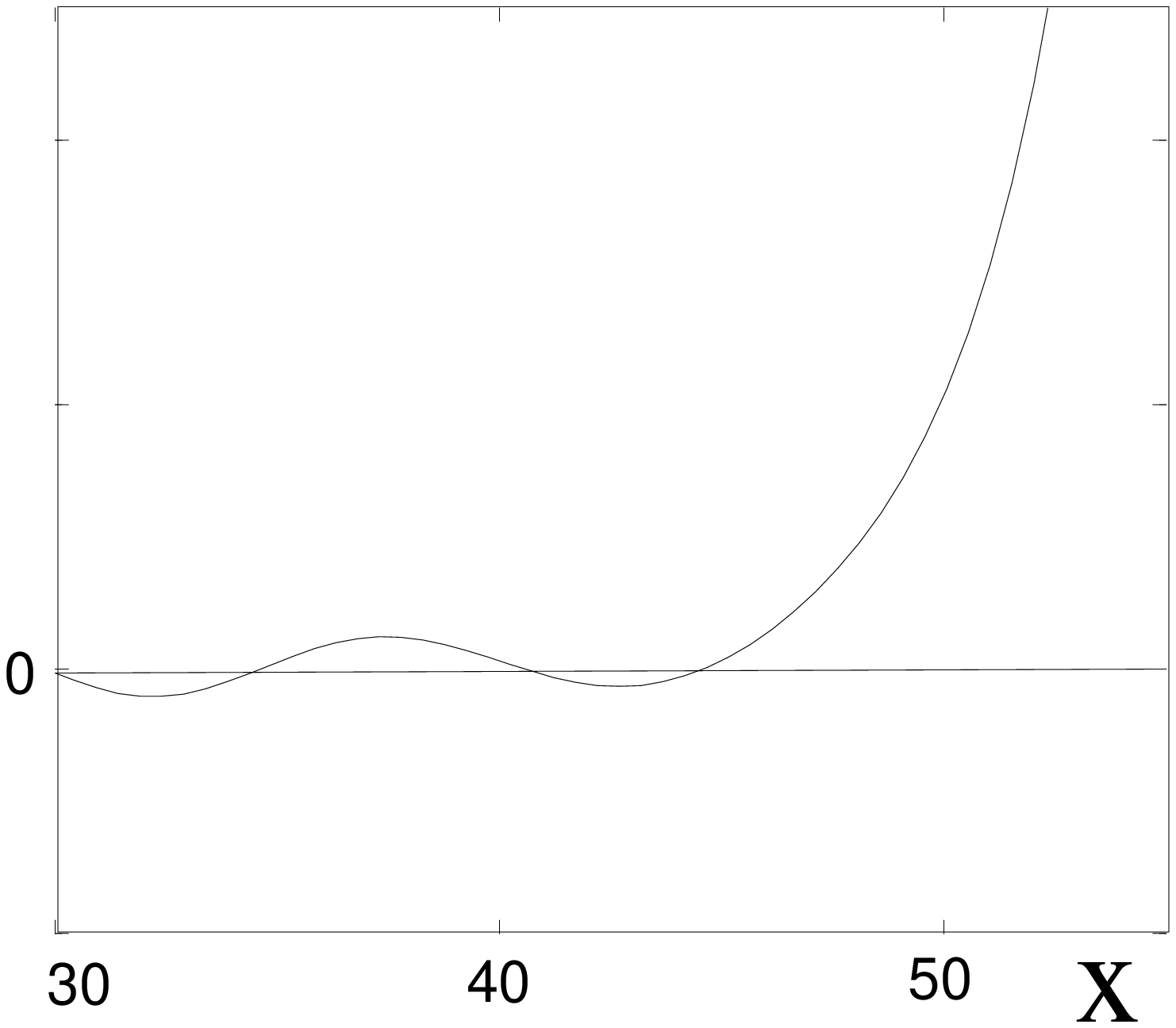,height=4cm,width=4cm}\hspace*{0.3cm}
 \epsfig{file=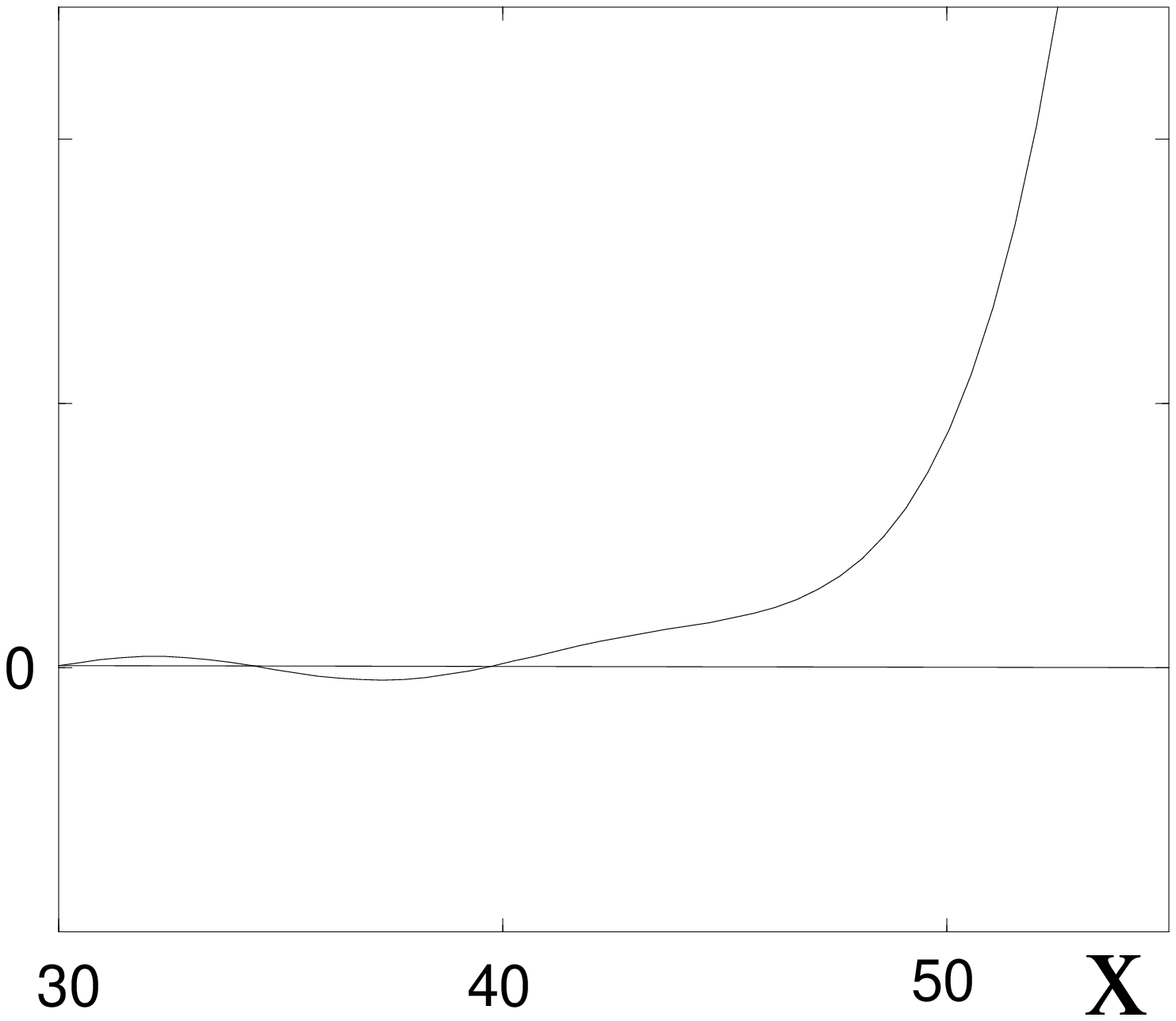,height=4cm,width=4cm}\\
\vspace*{-0.0cm}
Figure~1. Eeffects of the dressing terms.
\end{center}
\end{figure}
These three figures show the difference of the amplitude of radiations 
emitted during the time evolution for each initial wave. The initial
solitary wave is given by  
$u=v$ in the left figure, $u=v+\epsilon \phi^{(1)}$ in the central
figure, and $u=v+\epsilon \phi^{(1)}+\epsilon^2 
(\phi^{(2)}+\frac{1}{2}V_{\phi^{(1)}}\cdot \phi^{(1)})$ in the right figure. 
The core part is given by (\ref{eq:core}) for $\kappa=0.2$. 
These figures show that the generation of the radiation can be
suppressed by adding the higher order dressing terms into the core.

Now let us discuss the interaction of the two solitary waves.
In Figure 2, we show the result of the phase shift of the solitary wave with the parameter $\kappa_1 (>\kappa_2)$ for various values of
$\kappa_1-\kappa_2$.
Here we fix $\kappa_{1}+\kappa_{2}=0.5$ and 
$\epsilon=(\kappa_{1}^2+\kappa_{2}^2)/2 \simeq 0.07$.
The solid line is calculated from the formula
(\ref{eq:phaseshift}), and the broken line is obtained by the numerical simulation. The phase shift of the KdV soliton is also shown as the dotted line.
\begin{figure}[htbp]
\begin{center}
 \hspace*{2cm}$\frac{\Delta_{1}}{\epsilon^4 (\mu^{(2)}_{1})^2}$
\epsfig{file=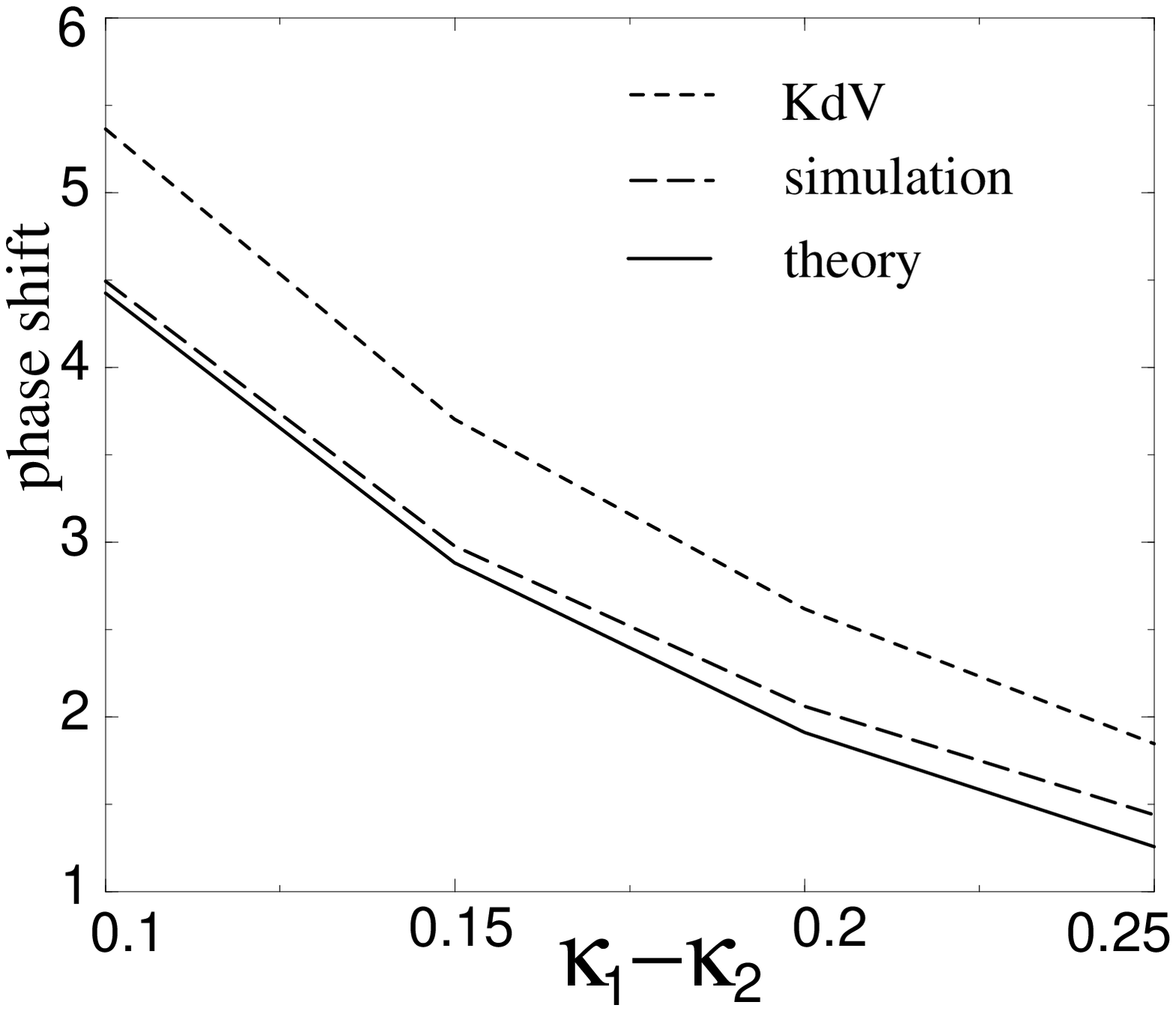,height=4.3cm,width=5.3cm}\hspace*{1cm} \epsfig{file=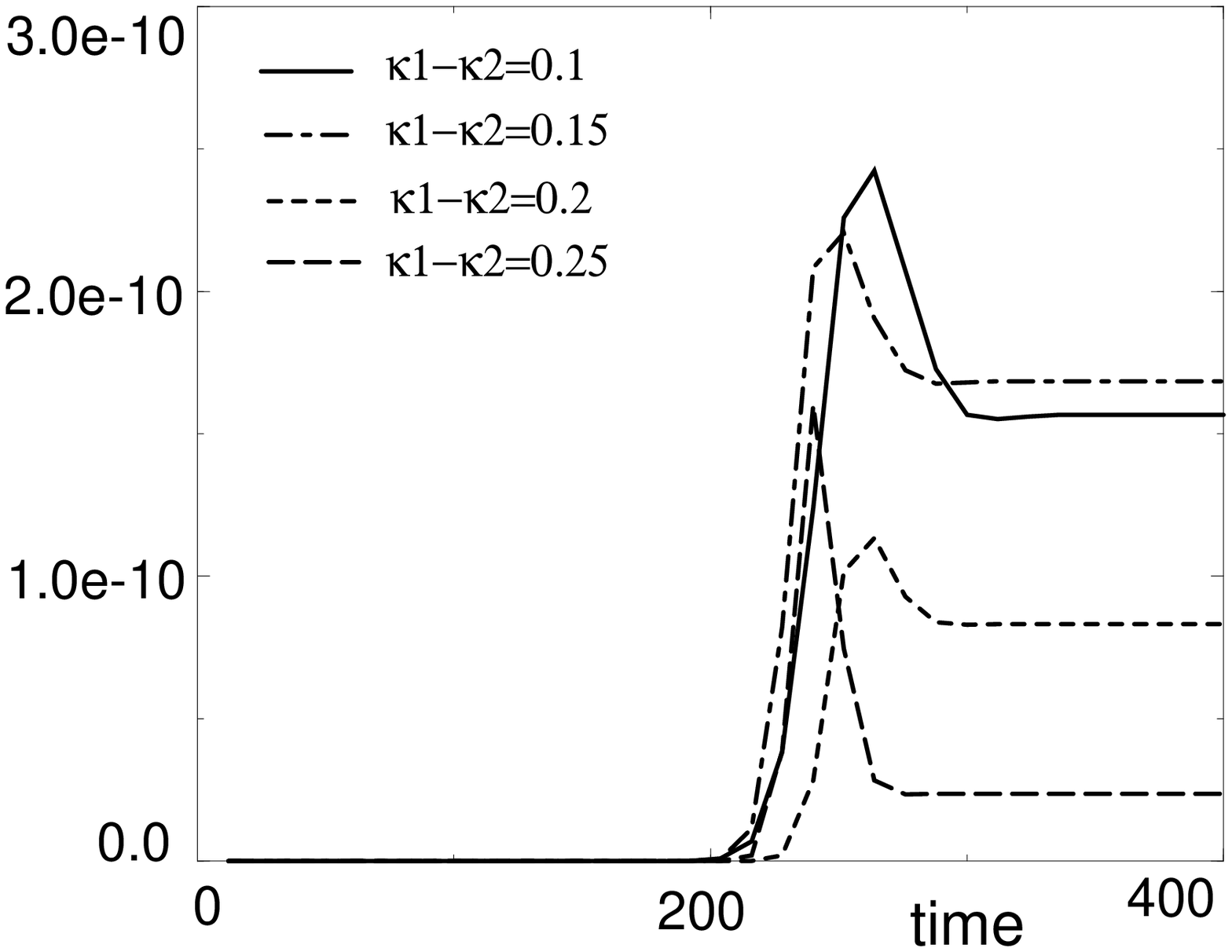,height=4.3cm,width=5.3cm} \\
\vspace*{0.3cm}
\hspace*{1.3cm}Figure~2. Phase shift. \hspace*{2.5cm}Figure~3. Time evolution of $\Delta_{1}(t)$.
\end{center}
\end{figure}
 As we can see that the phase shift formula 
gives a good agreement to the numerical results. The energy of the
radiation emitted after the interaction is caluculated from
(\ref{eq:deltam}) and is shown in Figure 3. This shows that the radiation energy is of order $10^{-5}$ which also agrees with the result in \cite{sattinger}. Thus the normal form theory provides an accurate description of the deviation from the KdV equation.

\subsection{Boussinesq equation}
The Boussinesq equation as an approximate equation for the shallow water waves is given by
\[\left\{\begin{array}{ll}
&\displaystyle{\eta_{T}+v_{X}+(\eta v)_{X}=0,} \\ &\displaystyle{v_{T}+\frac{1}{2}(v^2)_{X}+\eta_{X}-\frac{1}{3}v_{XXT}=0,}
\end{array}\right.\]
where $\eta$ and $v$ are the normalized variables which represent the amplitude and the velocity \cite{whitham}. Since this equation is truncated at the first order from the shallow water wave equation, the normal form may not provide a structure of the asymptotic integrability at the second order. However the Boussinesq equation itself has an interesting mathematical structure, such as the regularization at the higher dispersion regime, and it may be interesting to study its own asymptotic integrability. The normal form for the shallow water wave equation has been studied in \cite{kodama:88}.

Following the similar process as in the case of ion acoustic waves, we obtain the perturbed KdV equation up to order $\epsilon^2$,
\[\begin{array}{lll}
&\displaystyle{u_{t}+6uu_{x}+u_{3x}+\epsilon\left(\frac{3}{8}u_{5x}+\frac{3}{2}u_{3x}u+4u_{2x}{x}
-\frac{3}{4}u_{x}u^2\right)+\epsilon^2\left(\frac{5}{32}u_{7x}+\frac{11}{16}u_{5x}u
\right.}\\
&{}\\
&\displaystyle{\left.+\frac{99}{32}u_{4x}u_{x}+\frac{95}{16}u_{3x}u_{2x}-\frac{3}{16}u_{3x}u^2
-\frac{33}{16}u_{2x}u_{x}u-\frac{27}{32}u_{x}^3+\frac{3}{16}u_{x}u^3\right)=
O(\epsilon^3). }\end{array}\]

Due to the non-zero integrability condition
($\mu^{(2)}_{1}=3/2$), the obstacle $R^{(2)}_{1}$ does not
disappear in the second order correction. The generating functions in the Lie
transformation are given by 
\[ \left\{ \begin{array}{llll}
&\displaystyle{\alpha^{(1)}_{1}=\frac{13}{8},\quad\alpha^{(1)}_{2}=\frac{17}{24},
\quad\alpha^{(1)}_{3}=\frac{3}{4}, }\\
&{}\\
&\displaystyle{\alpha^{(2)}_{1}=\frac{2591}{14400},\quad\alpha^{(2)}_{2}=\frac{71}{100},
\quad\alpha^{(2)}_{3}=\frac{2747}{3600},} \\
&{}\\
&\displaystyle{\alpha^{(2)}_{4}=\frac{1583}{9600},\quad\alpha^{(2)}_{5}=\frac{17}{800},
\quad\alpha^{(2)}_{6}=\frac{571}{4800}. }
\end{array}\right. \]
Now let us consider the phase shift during the two solitary wave interactions
in the same way as the ion acoustic wave equation. The result is shown in
Figure 4. \\
\begin{figure}[htbp]
\begin{center}
 \hspace*{2cm}$\frac{\Delta_{1}}{\epsilon^4 {\mu^{(2)}_{1}}^2}$
 \epsfig{file=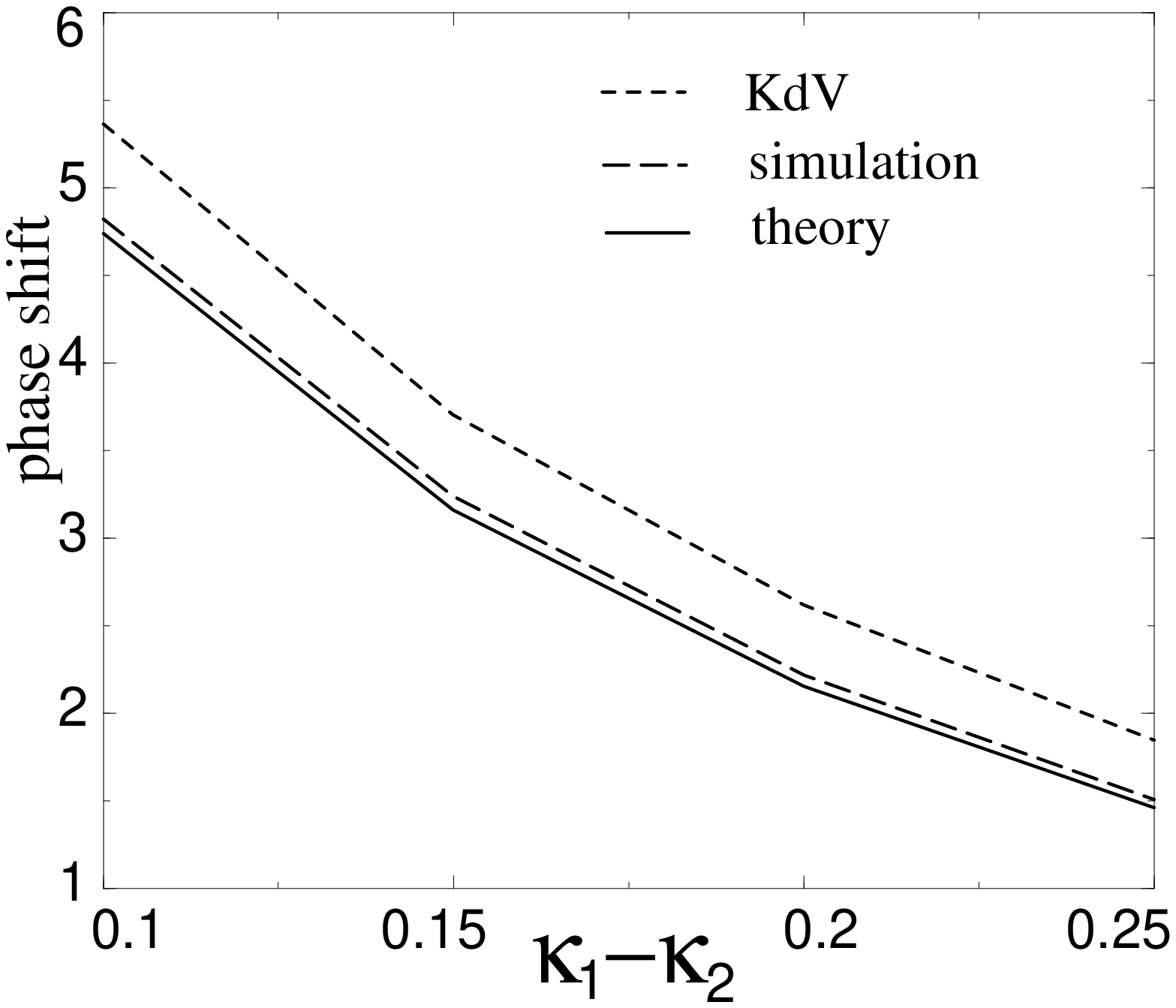,height=4.3cm,width=5.3cm}\hspace*{1cm} \epsfig{file=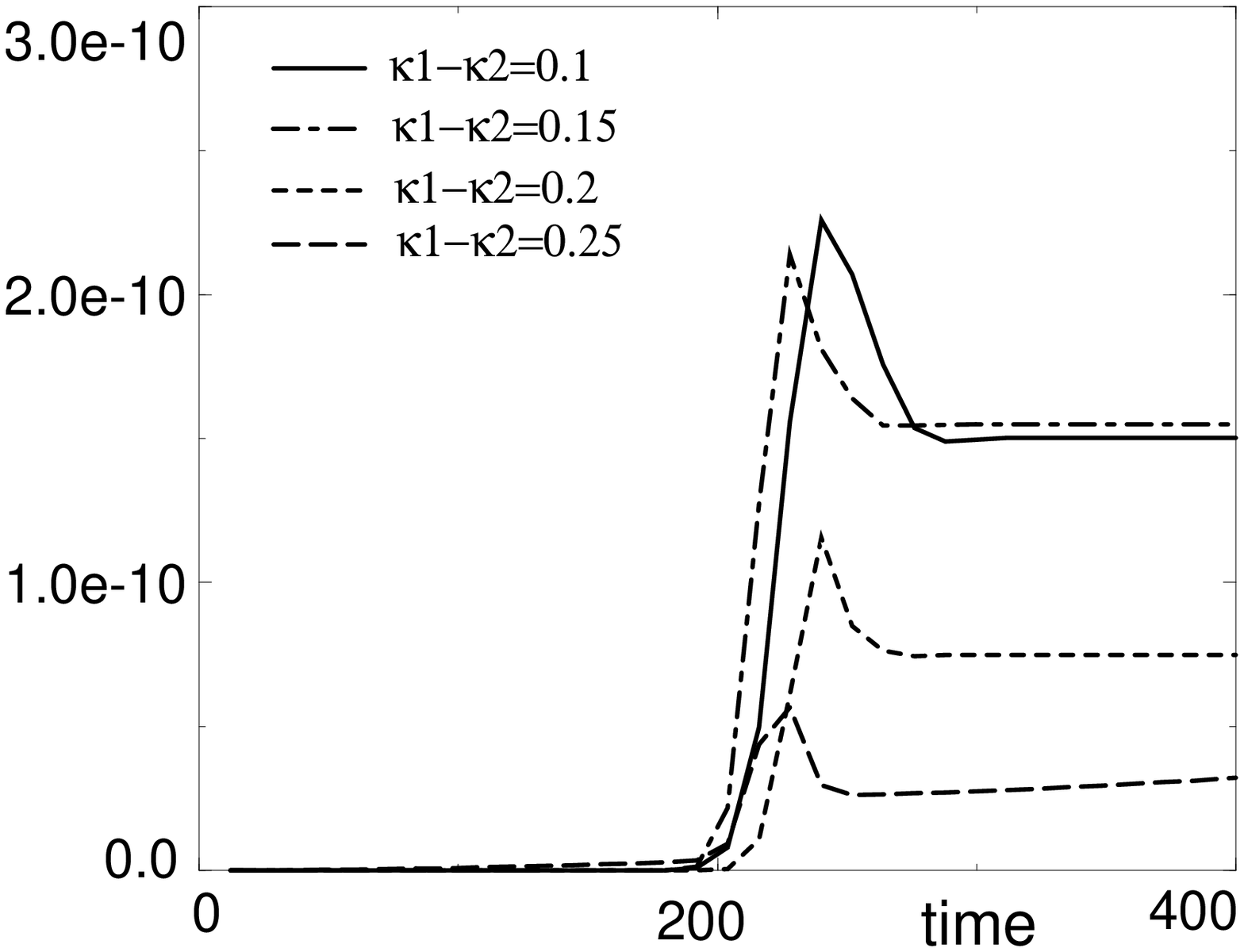,height=4.3cm,width=5.3cm} \\
\vspace*{0.3cm}
\hspace*{1.3cm}Figure~4. Phase shift. \hspace*{2.5cm}Figure~5. Time evolution of $\Delta_{1}(t)$.
\end{center}
\end{figure}
Again, the numerical data of the phase shift can be well explained by
the phase shift formula. The time evolution of $\Delta_{1}(t)$ is shown
in Figure 5. This figure shows that the energy of the radiation after the 
interaction is about the same as that for the ion acoustic wave equation.

\subsection{Regularized long wave equation}
Since the KdV equation is derived under the assumption of weak dispersion, it is not valid for the wave phenomena involving short waves.  
As a model of shallow water waves for a large range of wavelength, 
Benjamin et al in~\cite{bbm2, bbm1} proposed the equation
\[
w_{T}+w_{X}+6ww_{X}-w_{XXT}=0. \label{eq:bbm}
\]
This equation is called the regularized long wave equation (RLW) or the
BBM equation. It has been shown that
this equation has only three non-trivial independent conserved
quantities~\cite{bbmconservation} indicating its nonintegrability. The BBM equation admits a solitary wave solution,
\[
w=\frac{2\kappa^2}{1-4\kappa^2}{\rm sech}^2\left(\kappa X-\frac{\kappa}{1-4\kappa^2}T \right).
\]
The numerical simulations by Bona et al.~\cite{bbmnumerical} showed that the interaction of the solitary waves is inelastic and generates radiations. Their study also
found the shifts of the amplitudes of two solitary waves after the
collision. 
The normal form theory has been applied to this equation in~\cite{bbmnormal}. Here we review the work~\cite{bbmnormal}
and add the phase shift results. We also compare the BBM equation
with the ion acoustic wave equation and the Boussinesq equation. 

As in the previous cases, we introduce the scaled variables,
\[
x=\epsilon^{\frac{1}{2}}(X-T),\quad t=\epsilon^{\frac{3}{2}}T,\quad w=\epsilon u.
\]
which yields 
\[
(1-\epsilon D^2)u_{t}+6uu_{x}+u_{3x}=0. 
\]
Then inverting the operator in front of $u_t$, we obtain the perturbed
KdV  equation, 
\[
u_{t}+K_0^{(0)}(u)+\sum_{k=1}^{\infty}\epsilon^k D^{2k}K_0^{(0)}(u)=0. \]
Then the obstacle $R^{(2)}(u)$ appears with $\mu^{(2)}_{1}=40$. 
The comparison of the phase
shift between the numerical results and those by the formula
(\ref{eq:phaseshift}) is shown in Figure 6. 
\begin{figure}[htbp]
\begin{center}
 \epsfig{file=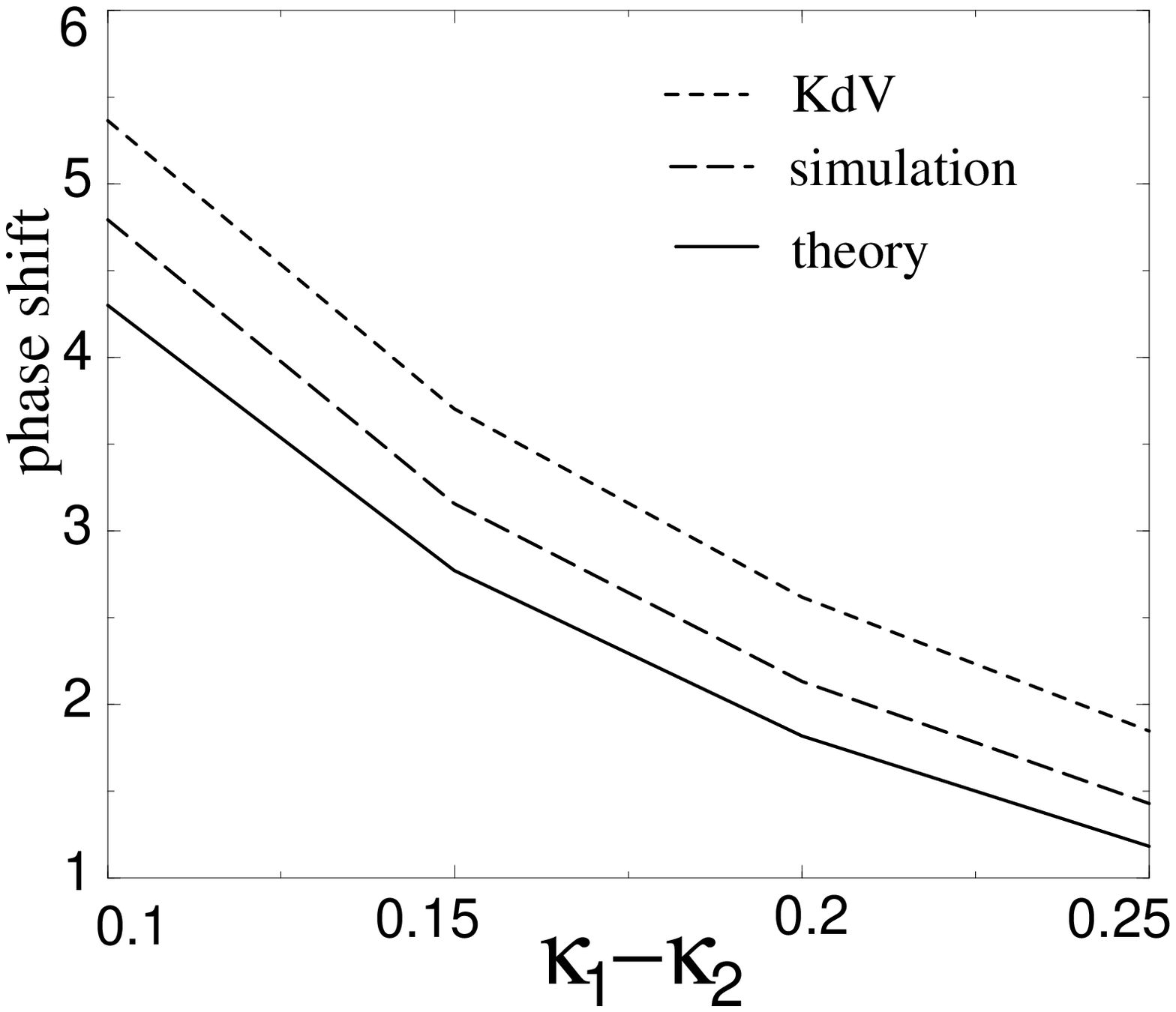,height=4.3cm,width=5.3cm} \\
\vspace*{0.3cm}
\hspace*{0.7cm}Figure 6. phase shift.
\end{center}
\end{figure}
Because of the large value of $\mu_1^{(2)}$, the agreement between the numerical results and
the results from (\ref{eq:phaseshift}) is poor. In the computation, the value of $\epsilon^2\mu_1^{(2)}$ is about of order one, and thus
one needs to consider much higher corrections to get a better agreement.
However the numerical observation of the radiation agrees with the formula (\ref{eq:deltam}) which is of order $10^{-2}\sim 10^{-3}$, and
the shifts in the parameters observed in \cite{bbmnumerical} also agrees with the result from the normal form theory.

\subsection{7th order Hirota KdV equation}
Here we consider 
\begin{equation}
\label{eq:hirotav}
\begin{array}{ll}
& w_{T}+w_{7X}+28w w_{5X}+28 w_{X}w_{4X}+70w_{2X}w_{3X} \\ 
&{}\\
&\quad +210w^2w_{3X}+420ww_Xw_{2X}+420w^3w_x=0,
\end{array}
 \end{equation}
which has a Hirota bilinear form,
\[
D_{X}(D_{X}^7+D_{T})\tau \cdot \tau=0 \label{eq:hirotatau},
\]
where $D_{X}$ and $D_{T}$ mean the Hirota derivative, and $w$ is given by $w=2D^2\ln\tau$.  
If the order of the derivative is either {\it three} or {\it five}, the equation become the KdV
equation or the Sawada-Kotera equation which are both integrable.
However (\ref{eq:hirotatau}) is known to be nonintegrable \cite{newell:85}, but it admits both 
 one and two solitary wave solutions in the same form as the KdV solitons except their time evolution. 
Since there is an exact two-solitary wave solution, several conditions
$\mu_k^{(n)}=0$ should be satisfied. 
Now the question is to determine the order in which the condition
$\mu_k^{(n)}=0$ breaks.

In order to apply the normal form theory, we first introduce
\[
w=\epsilon u+c,\quad t=210 \epsilon^{3/2} c^2 T,\quad  x=\epsilon^{1/2} (X-420 c^3 T),
\]
which puts
(\ref{eq:hirotav}) in the form (\ref{eq:pkdv}) of the perturbed KdV equation
\[\begin{array}{lll}
&\displaystyle{u_{t}+K_0^{(0)}(u)+\epsilon \frac{2}{15 c}(u_{5x}+15 u_{3x}u + 15 u_{2x} u_{x}+45 u_{x}u^2)  + \frac{\epsilon^2}{210c^2}(u_{7x}+28 u_{5x}u} \\
&{}\\
&\quad +28 u_{4x}u_{x}+70u_{3x}u_{2x}+210u_{3x}u^2 +420u_{2x}u_{x}u+420u_{x}u^3)=0,
\end{array}\]
where $c$ is an arbitrary non-zero constant. 
The direct calculation shows 
\[\left\{\begin{array}{lll}
&\mu^{(2)}_{1}=0, \\
&\mu^{(3)}_{i}=0,\quad &i=1,2,3, \\
&\mu^{(4)}_{i}\ne 0,\quad &i=1,2,\cdots,7. 
\end{array}\right.\]
Thus the 7th order Hirota KdV equation passes the asymptotic integrability conditions 
not only at order $\epsilon^2$ but also at order $\epsilon^3$, and 
the first obstacles appear at order $\epsilon^4$. 
Then  the normal form up to 
order $\epsilon^4$ of the 7th order Hirota KdV equation takes the form,
\begin{equation}
\label{eq:4thnormal}
v_{t}+K^{(0)}_0+\epsilon a^{(1)}_{1}K_{0}^{(1)}+\epsilon^2 a^{(2)}_{1}K_{0}^{(2)}
+\epsilon^3 a^{(3)}_{1}K_{0}^{(3)}+\epsilon^4 (a^{(4)}_{1}K_{0}^{(4)}+R^{(4)})=O(\epsilon^5), 
\end{equation}
where $R^{(4)}$ consists of seven obstacles. 
Since the 7th order Hirota KdV equation admits the two-soliton 
solution, the kernel of $R^{(4)}$ should include not only one-soliton solution but also two-soliton solution.
Thus we have
\begin{Corollary} 
If the normal form of a perturbed KdV equation has the form (\ref{eq:4thnormal}), 
then there exists two-soliton solution up to order 
$\epsilon^4$. Equivalently,  
if $\mu^{(2)}_{1}$, $\mu^{(3)}_{1}$, $\mu^{(3)}_{2}$, and $\mu^{(3)}_{3}$ are 
all zero, then the pertubed KdV equation can admit a two-soliton solution up to order $\epsilon^4$.
\end{Corollary}

\appendix
\section*{Appendix}
\label{appendixa}

Coefficients of the generating function $\phi^{(2)}$:
\[\begin{array}{lll}
\alpha^{(2)}_{1}=&-\frac{100}{9}(a^{(1)}_{1})^2-\frac{10}{27}a^{(1)}_{1}a^{(1)}_{2}
-\frac{5}{108}(a^{(1)}_{2})^2-\frac{1}{54}a^{(1)}_{2}a^{(1)}_{4}+\frac{1}{36}(a^{(1)}_{4})^2\\
&{}\\
&+\frac{112}{9}a^{(2)}_{1}+\frac{2}{9}a^{(2)}_{2}+\frac{1}{9}a^{(2)}_{4}
-\frac{1}{6}a^{(2)}_{7}-\frac{4}{225}\mu^{(2)}_{1}, \\
& \nonumber \\
\alpha^{(2)}_{2}=&-\frac{175}{9}(a^{(1)}_{1})^2+\frac{10}{9}a^{(1)}_{1}a^{(1)}_{2}
+\frac{5}{18}a^{(1)}_{1}a^{(1)}_{3}-\frac{1}{36}a^{(1)}_{2}a^{(1)}_{3}
-\frac{5}{18}a^{(1)}_{1}a^{(1)}_{4}-\frac{1}{18}a^{(1)}_{2}a^{(1)}_{4}\\
&{}\\
&-\frac{1}{36}a^{(1)}_{3}a^{(1)}_{4}+\frac{1}{18}(a^{(1)}_{4})^2 +\frac{56}{3}a^{(2)}_{1}-\frac{1}{2}a^{(2)}_{2}+\frac{1}{12}a^{(2)}_{6}
-\frac{1}{4}a^{(2)}_{7}-\frac{2}{75}\mu^{(2)}_{1}, \\
&& \nonumber \\
\alpha^{(2)}_{3}=&-\frac{25}{3}(a^{(1)}_{1})^2-\frac{85}{108}a^{(1)}_{1}a^{(1)}_{2}
+\frac{1}{108}(a^{(1)}_{2})^2+\frac{5}{18}a^{(1)}_{1}a^{(1)}_{3}
+\frac{1}{72}a^{(1)}_{2}a^{(1)}_{3} -\frac{5}{36}a^{(1)}_{1}a^{(1)}_{4} \\
&{}\\
&-\frac{1}{216}a^{(1)}_{2}a^{(1)}_{4}-\frac{1}{24}a^{(1)}_{3}a^{(1)}_{4}
+\frac{1}{24}(a^{(1)}_{4})^2+\frac{91}{9}a^{(2)}_{1}+\frac{7}{18}a^{(2)}_{2}
-\frac{1}{18}a^{(2)}_{4}-\frac{1}{6}a^{(2)}_{7}  \\
&{}\\
&+\frac{1}{6}a^{(2)}_{8}-\frac{13}{900}\mu^{(2)}_{1}, \\
&& \nonumber \\
\alpha^{(2)}_{4}=&-\frac{25}{36}(a^{(1)}_{1})^2+\frac{5}{48}a^{(1)}_{1}a^{(1)}_{2}
-\frac{1}{144}a^{(1)}_{2}a^{(1)}_{3}-\frac{5}{72}a^{(1)}_{1}a^{(1)}_{4}
-\frac{1}{288}a^{(1)}_{2}a^{(1)}_{4}-\frac{1}{288}a^{(1)}_{3}a^{(1)}_{4} \\
&{}\\
&+\frac{1}{144}(a^{(1)}_{4})^2+\frac{7}{6}a^{(2)}_{1}-\frac{5}{48}a^{(2)}_{2}
+\frac{1}{24}a^{(2)}_{3}+\frac{1}{96}a^{(2)}_{6}-\frac{1}{32}a^{(2)}_{7}
-\frac{7}{2400}\mu^{(2)}_{1}, \\
&& \nonumber \\
\alpha^{(2)}_{5}=&-\frac{50}{9}(a^{(1)}_{1})^2+\frac{5}{9}a^{(1)}_{1}a^{(1)}_{2}
+\frac{14}{3}a^{(2)}_{1}-\frac{1}{3}a^{(2)}_{2}-\frac{1}{150}\mu^{(2)}_{1}, \\
&& \nonumber \\
\alpha^{(2)}_{6}=&\frac{50}{3}(a^{(1)}_{1})^2-\frac{85}{18}a^{(1)}_{1}a^{(1)}_{2}
+\frac{5}{36}(a^{(1)}_{2})^2+\frac{5}{18}a^{(1)}_{1}a^{(1)}_{4}
+\frac{1}{36}a^{(1)}_{2}a^{(1)}_{4}-\frac{28}{3}a^{(2)}_{1} \\
&{}\\
&+\frac{7}{3}a^{(2)}_{2}-\frac{1}{3}a^{(2)}_{4}+\frac{1}{75}\mu^{(2)}_{1}.\\
\end{array}\]

The values $\mu_k^{(3)}$ for $k=1,2,3$:

\[\begin{array}{lll}  \mu^{(3)}_{1}=&\frac{280}{3}a^{(3)}_{2}-\frac{280}{3}a^{(3)}_{3}
+60a^{(3)}_{4}-20a^{(3)}_{6}-\frac{5}{3}a^{(3)}_{7}-a^{(3)}_{8}
+8a^{(3)}_{9}-a^{(3)}_{10}-4a^{(3)}_{11}\\
&{}\\
&+\frac{2}{3}a^{(3)}_{12}-a^{(3)}_{13}-2a^{(3)}_{15} +\frac{800}{3}(a^{(1)}_{1})^3+\frac{130}{9}(a^{(1)}_{1})^2a^{(1)}_{2}
+\frac{62}{27}a^{(1)}_{1}(a^{(1)}_{2})^2-\frac{2}{27}(a^{(1)}_{2})^3\\
&{}\\
&-\frac{400}{9}(a^{(1)}_{1})^2a^{(1)}_{3}-\frac{175}{27}a^{(1)}_{1}a^{(1)}_{2}a^{(1)}_{3} +\frac{1}{9}(a^{(1)}_{2})^2a^{(1)}_{3}+\frac{34}{9}a^{(1)}_{1}(a^{(1)}_{3})^2
+\frac{1}{9}a^{(1)}_{2}(a^{(1)}_{3})^2\\
&{}\\
&-\frac{2}{27}(a^{(1)}_{3})^3
+\frac{20}{3}(a^{(1)}_{1})^2a^{(1)}_{4}-\frac{7}{27}a^{(1)}_{1}a^{(1)}_{2}a^{(1)}_{4}  -\frac{4}{27}a^{(1)}_{1}a^{(1)}_{3}a^{(1)}_{4}-\frac{560}{3}a^{(1)}_{1}a^{(2)}_{1}\\
&{}\\
&-\frac{560}{9}a^{(1)}_{2}a^{(2)}_{1}+\frac{280}{9}a^{(1)}_{3}a^{(2)}_{1}
-\frac{230}{3}a^{(1)}_{1}a^{(2)}_{2}+\frac{10}{9}a^{(1)}_{2}a^{(2)}_{2} +\frac{55}{9}a^{(1)}_{3}a^{(2)}_{2}+\frac{340}{3}a^{(1)}_{1}a^{(2)}_{3}\\
&{}\\
&-\frac{10}{9}a^{(1)}_{2}a^{(2)}_{3}-\frac{20}{3}a^{(1)}_{3}a^{(2)}_{3}
+\frac{5}{9}a^{(1)}_{4}a^{(2)}_{3}-\frac{64}{9}a^{(1)}_{1}a^{(2)}_{4}
+\frac{1}{3}a^{(1)}_{2}a^{(2)}_{4}-\frac{1}{3}a^{(1)}_{3}a^{(2)}_{4}\\
&{}\\
&-\frac{136}{3}a^{(1)}_{1}a^{(2)}_{5}
+2a^{(1)}_{2}a^{(2)}_{5}+2a^{(1)}_{3}a^{(2)}_{5}-\frac{1}{3}a^{(1)}_{4}a^{(2)}_{5}
+\frac{28}{9}a^{(1)}_{1}a^{(2)}_{6}-\frac{1}{6}a^{(1)}_{2}a^{(2)}_{6}\\
&{}\\
&+a^{(1)}_{1}a^{(2)}_{7}-12a^{(1)}_{1}a^{(2)}_{8}-\frac{1}{3}a^{(1)}_{2}a^{(2)}_{8}
+\frac{2}{3}a^{(1)}_{3}a^{(2)}_{8}+\frac{1}{3}\mu^{(2)}_{1},
 \end{array}\]
 
\[\begin{array}{lll}
\mu^{(3)}_{2} =& -4200a^{(3)}_{1}+770a^{(3)}_{2}-\frac{280}{3}a^{(3)}_{3}-40a^{(3)}_{4}
-\frac{130}{3}a^{(3)}_{5}+40a^{(3)}_{6}+\frac{10}{3}a^{(3)}_{7}-8a^{(3)}_{8}\\
&{}\\
&-6a^{(3)}_{9}+7a^{(3)}_{10}+8a^{(3)}_{11}-\frac{4}{3}a^{(3)}_{12}+2a^{(3)}_{13}
+9a^{(3)}_{15}-\frac{95000}{9}(a^{(1)}_{1})^3+\frac{73450}{27}(a^{(1)}_{1})^2a^{(1)}_{2}\\
&{}\\
&-\frac{325}{3}a^{(1)}_{1}(a^{(1)}_{2})^2+\frac{14}{27}(a^{(1)}_{2})^3
+\frac{800}{9}(a^{(1)}_{1})^2a^{(1)}_{3} -\frac{1720}{27}a^{(1)}_{1}a^{(1)}_{2}a^{(1)}_{3}\\
&{}\\
&+\frac{8}{9}(a^{(1)}_{2})^2a^{(1)}_{3}
+\frac{70}{9}a^{(1)}_{1}(a^{(1)}_{3})^2-\frac{2}{9}a^{(1)}_{2}(a^{(1)}_{3})^2 +\frac{4}{27}(a^{(1)}_{3})^3-\frac{500}{3}(a^{(1)}_{1})^2a^{(1)}_{4} \\
&{}\\
&+10a^{(1)}_{1}a^{(1)}_{2}a^{(1)}_{4}+\frac{65}{27}a^{(1)}_{1}a^{(1)}_{3}a^{(1)}_{4}
+\frac{123200}{9}a^{(1)}_{1}a^{(2)}_{1}-\frac{12250}{9}a^{(1)}_{2}a^{(2)}_{1}\\ 
&{}\\
&-\frac{560}{9}a^{(1)}_{3}a^{(2)}_{1}+\frac{280}{3}a^{(1)}_{4}a^{(2)}_{1}
-\frac{15950}{9}a^{(1)}_{1}a^{(2)}_{2}+\frac{685}{9}a^{(1)}_{2}a^{(2)}_{2}
+\frac{310}{9}a^{(1)}_{3}a^{(2)}_{2}\\
&{}\\
&-\frac{55}{9}a^{(1)}_{4}a^{(2)}_{2}+\frac{700}{3}a^{(1)}_{1}a^{(2)}_{3}
-\frac{40}{9}a^{(1)}_{2}a^{(2)}_{3}
-\frac{10}{9}a^{(1)}_{4}a^{(2)}_{3}+\frac{310}{3}a^{(1)}_{1}a^{(2)}_{4}\\
&{}\\
&-\frac{7}{3}a^{(1)}_{2}a^{(2)}_{4}-a^{(1)}_{3}a^{(2)}_{4}
-\frac{280}{3}a^{(1)}_{1}a^{(2)}_{5}+\frac{28}{3}a^{(1)}_{2}a^{(2)}_{5} -4a^{(1)}_{3}a^{(2)}_{5}+\frac{2}{3}a^{(1)}_{4}a^{(2)}_{5}\\
&{}\\
&+\frac{115}{9}a^{(1)}_{1}a^{(2)}_{6}-\frac{1}{2}a^{(1)}_{2}a^{(2)}_{6}
-\frac{25}{3}a^{(1)}_{1}a^{(2)}_{7}-\frac{80}{3}a^{(1)}_{1}a^{(2)}_{8}
+\frac{7}{3}a^{(1)}_{2}a^{(2)}_{8}\\
&{}\\
&-\frac{4}{3}a^{(1)}_{3}a^{(2)}_{8}-\frac{7}{3}\mu^{(2)}_{1}, 
 \end{array}\]

\[\begin{array}{lll}  \mu^{(3)}_{3}=&560a^{(3)}_{2}-\frac{1400}{3}a^{(3)}_{3}+300a^{(3)}_{4}
-\frac{170}{3}a^{(3)}_{5}-100a^{(3)}_{6}+\frac{5}{3}a^{(3)}_{7}-9a^{(3)}_{8}
+40a^{(3)}_{9}\\
&{}\\
&-a^{(3)}_{10}-20a^{(3)}_{11}+\frac{19}{3}a^{(3)}_{12}
-9a^{(3)}_{13}-2a^{(3)}_{14}-10a^{(3)}_{15}
+\frac{32000}{9}(a^{(1)}_{1})^3\\
&{}\\
&-\frac{1600}{27}(a^{(1)}_{1})^2a^{(1)}_{2}-\frac{2270}{27}a^{(1)}_{1}(a^{(1)}_{2})^2
-\frac{2800}{9}(a^{(1)}_{1})^2a^{(1)}_{3}
+\frac{545}{27}a^{(1)}_{1}a^{(1)}_{2}a^{(1)}_{3}\\
&{}\\
&+\frac{86}{27}(a^{(1)}_{2})^2a^{(1)}_{3}+\frac{40}{3}a^{(1)}_{1}(a^{(1)}_{3})^2
-\frac{10}{9}a^{(1)}_{2}(a^{(1)}_{3})^2-\frac{4}{27}(a^{(1)}_{3})^3
-\frac{400}{9}(a^{(1)}_{1})^2a^{(1)}_{4}\\
&{}\\
&+30a^{(1)}_{1}a^{(1)}_{2}a^{(1)}_{4}-\frac{1}{27}(a^{(1)}_{2})^2a^{(1)}_{4}
-\frac{155}{27}a^{(1)}_{1}a^{(1)}_{3}a^{(1)}_{4}
-\frac{19}{27}a^{(1)}_{2}a^{(1)}_{3}a^{(1)}_{4}\\
&{}\\
&+\frac{2}{9}(a^{(1)}_{3})^2a^{(1)}_{4}
-\frac{10}{9}a^{(1)}_{1}(a^{(1)}_{4})^2-\frac{22400}{9}a^{(1)}_{1}a^{(2)}_{1}
-560a^{(1)}_{2}a^{(2)}_{1}+\frac{1960}{9}a^{(1)}_{3}a^{(2)}_{1}  \\
&{}\\
&+\frac{280}{3}a^{(1)}_{4}a^{(2)}_{1}-\frac{1600}{9}a^{(1)}_{1}a^{(2)}_{2}
+90a^{(1)}_{2}a^{(2)}_{2}+\frac{35}{3}a^{(1)}_{3}a^{(2)}_{2}
-\frac{200}{9}a^{(1)}_{4}a^{(2)}_{2}\\
&{}\\
&+400a^{(1)}_{1}a^{(2)}_{3} -\frac{200}{9}a^{(1)}_{2}a^{(2)}_{3}-\frac{80}{3}a^{(1)}_{3}a^{(2)}_{3}
+\frac{85}{9}a^{(1)}_{4}a^{(2)}_{3}+\frac{50}{3}a^{(1)}_{1}a^{(2)}_{4}\\
&{}\\
&-\frac{1}{9}a^{(1)}_{2}a^{(2)}_{4}-\frac{37}{9}a^{(1)}_{3}a^{(2)}_{4} +\frac{1}{3}a^{(1)}_{4}a^{(2)}_{4}-160a^{(1)}_{1}a^{(2)}_{5}
+\frac{50}{3}a^{(1)}_{2}a^{(2)}_{5}\\
&{}\\
&+\frac{22}{3}a^{(1)}_{3}a^{(2)}_{5}
-\frac{13}{3}a^{(1)}_{4}a^{(2)}_{5}-\frac{85}{9}a^{(1)}_{1}a^{(2)}_{6}
-\frac{7}{2}a^{(1)}_{2}a^{(2)}_{6}+\frac{5}{3}a^{(1)}_{3}a^{(2)}_{6}\\
&{}\\
&+\frac{2}{3}a^{(1)}_{4}a^{(2)}_{6}
+\frac{10}{3}a^{(1)}_{1}a^{(2)}_{7}-\frac{1}{3}a^{(1)}_{2}a^{(2)}_{7}
+\frac{1}{6}a^{(1)}_{3}a^{(2)}_{7}+\frac{20}{3}a^{(1)}_{1}a^{(2)}_{8}\\
&{}\\
&+a^{(1)}_{2}a^{(2)}_{8} +\frac{2}{3}a^{(1)}_{3}a^{(2)}_{8}-\frac{4}{3}a^{(1)}_{4}a^{(2)}_{8} +\frac{7}{3}\mu^{(2)}_{1}.
\end{array}\]

\end{document}